\documentclass[5p, twocolumn]{elsarticle}

\usepackage{lineno,hyperref}
\usepackage{pifont}
\usepackage{natbib}
\usepackage{geometry}
\usepackage{graphicx}
\usepackage{amsmath}
\usepackage{color}
\modulolinenumbers[5]
\journal{Acta Materialia}

\usepackage{amssymb}

\newcommand{\R}{\mathbb{R}}
\renewcommand{\d}[1]{\mathinner{d#1}}
\newcommand{\fn}[2]{\mathinner{#1\mathopen{\left(#2\right)}}}
\newcommand{\vect}[1]{\mathbf{#1}}

\newcommand{\abs}[1]{\left\vert #1 \right\vert}
\newcommand{\ceil}[1]{\lceil #1 \rceil}

\newcommand{\vv}[1]{\fn{\sigma_V ^2}{#1}}
\newcommand{\spD}[1]{\fn{\tilde{\chi}_{_V}}{#1}}

\begin{document}

\begin{frontmatter}

\title{New Tessellation-Based Procedure to Design Perfectly Hyperuniform Disordered Dispersions for Materials Discovery}

\author[phys]{J. Kim}

\author[phys,chem,prism,appmath]{S. Torquato\corref{mycorrespondingauthor}}
\cortext[mycorrespondingauthor]{Corresponding author}
\ead{torquato@electron.princeton.edu}
\ead[url]{http://chemlabs.princeton.edu/torquato/}
\address[phys]{Department of Physics, Princeton University, Princeton, New Jersey 08544, USA}
\address[chem]{Department of Chemistry, Princeton University, Princeton, New Jersey 08544, USA}
\address[prism]{Princeton Institute for the Science and Technology of Materials, Princeton University, Princeton, New Jersey 08544, USA}
\address[appmath]{Program in Applied and Computational Mathematics, Princeton University, Princeton, New Jersey 08544, USA}

\begin{abstract}
Disordered hyperuniform dispersions are exotic amorphous two-phase materials characterized by an anomalous suppression of long-wavelength volume-fraction fluctuations, which endows them with novel physical properties. 
While such unusual materials have received considerable attention, a stumbling block has been an inability to create large samples that are truly hyperuniform due to current computational and experimental limitations. 
To overcome such barriers, we introduce a new and simple construction procedure that guarantees perfect hyperuniformity for very large sample sizes.
This methodology involves tessellating space into cells and then inserting a particle into each cell such that the local-cell particle packing fractions are identical to the global packing fraction.
We analytically prove that such dispersions are perfectly hyperuniform in the infinite-sample-size limit.
Our methodology enables a remarkable mapping that converts a very large nonhyperuniform disordered dispersion into a \textit{perfectly} hyperuniform one, which we numerically demonstrate in two and three dimensions. 
A similar analysis also establishes the hyperuniformity of the famous Hashin-Shtrikman multiscale dispersions, which possess optimal transport and elastic properties.
Our hyperuniform designs can be readily fabricated using modern photolithographic and 3D printing technologies.
The exploration of the enormous class of hyperuniform dispersions that can be designed and tuned by our tessellation-based methodology paves the way for accelerating the discovery of novel hyperuniform materials.

\end{abstract}

\begin{keyword}
Disordered hyperuniformity\sep 
Heterogeneous materials\sep 
Hashin-Shtrikman coated-spheres model

\end{keyword}

\end{frontmatter}

\section{Introduction}\label{sec:introduction}

Disordered hyperuniform materials \cite{Torquato2003_hyper, Zachary2009a, Torquato2018_review} are exotic amorphous states of matter that are like crystals in the manner in which their large-scale density fluctuations are anomalously suppressed and yet behave like liquids or glasses in that they are statistically isotropic without any Bragg peaks.
Recent results offer glimpses into the remarkable physical properties that such unusual correlated disordered materials can possess, including complete isotropic photonic/phononic band gaps \cite{Florescu2009, Scheffold2017, Lopez2018},  nearly optimal transport and mechanical properties \cite{Zhang2016, Chen2017, Thien2016},  and dense but transparent materials \cite{Leseur2016}.

The hyperuniformity concept was first introduced in the context of point patterns \cite{Torquato2003_hyper} and later extended to two-phase systems \cite{Zachary2009a, Torquato2018_review},  such as composites, colloidal suspensions, and polymer blends \cite{Brinker_sol-gel, Sahimi_HM1, Torquato_RHM, Patel2016, Mickel2013}.
A hyperuniform two-phase material is one in which the local volume-fraction variance $\vv{R}$ inside a spherical observation window of radius $R$ decays faster than $R^{-d}$ for large $R$, which is the scaling for typical disordered systems.
In the strongest form of hyperuniformity, called ``class I,'' the large-$R$ scaling of the variance is $\vv{R}\sim R^{-(d+1)}$ \cite{Zachary2009a,Torquato2018_review}.
Equivalently, its spectral density function $\spD{\vect{k}}$, obtainable from scattering experiments \cite{Debye1949}, vanishes as the wavenumber $\abs{\vect{k}}$ tends to zero \cite{Zachary2009a, Torquato2018_review},  and hence hyperuniform systems encompass all periodic and special disordered systems.

To date, a variety of disordered hyperuniform systems have been identified, including classical equilibrium systems \cite{Uche2004, Torquato2015_stealthy, Zhang2015, Dyson1962, Jancovici1981, Duyu2018_2},  quantum systems \cite{Feynman1956, Torquato2008, Scardicchio2009},  maximally random jammed packings \cite{Donev2005,Kurita2011, Jiao2011},  
non-equilibrium critical states \cite{Hexner2015,Weijs2015}, non-equilibrium dynamical systems \cite{Garcia-Millan2018}, random speckle patterns \cite{Battista2018}, number theory \cite{Dyson1962,Torquato2018_prime2, Montgomery1973, Sodin2006}, and biological systems \cite{Noh2010, Jiao2014_chickenEyes, Mayer2015, Kwon2017}; see also a recent review \cite{Torquato2018_review} and references therein.
While some of these systems \cite{Uche2004, Torquato2015_stealthy, Zhang2015, Dyson1962, Jancovici1981, Duyu2018_2, Feynman1956, Torquato2008, Scardicchio2009, Montgomery1973, Sodin2006} are proved to be perfectly hyperuniform in the infinite-sample-size limit, others are effectively hyperuniform, i.e., $\spD{0}$ is not exactly zero but small compared to the peak value of the spectral density \cite{Torquato2018_review, Atkinson2016}. 
Although effective hyperuniformity is sufficient to exhibit desirable physical properties, it has been shown that driving systems into perfect hyperuniformity will further improve their performance \cite{Hejna2013, Florescu2010}.

Importantly, hyperuniformity is a property of infinitely large systems, which implies that one needs an arbitrarily large sample to ascertain whether the system is perfectly hyperuniform.
However, current experimental \cite{Kurita2011,Weijs2015,  Ricouvier2017, Zito2015} and computational \cite{Duyu2018_2, Atkinson2016, Zachary2011_3, Lomba2017, TJ_algorithm, LS_algorithm} methods are limited in their capacity to create perfectly hyperuniform materials -- increasing sample size comes at the expense of perfect hyperuniformity leading to effective hyperuniformity at best.  
For instance, while the collective-coordinate optimization technique \cite{Zhang2016, Chen2017, Uche2004} ensures the generation of perfectly hyperuniform disordered many-particle systems, even at finite wavelengths, its computational cost grows rapidly with sample size.
Furthermore, hyperuniformity  can be degraded or destroyed (even if by a small amount) due to the inevitable presence of imperfections in otherwise perfectly hyperuniform systems \cite{Atkinson2016, Jaeuk2018}.

Hence, there is a great need to devise systematic procedures to construct extremely large realizations of perfectly hyperuniform disordered two-phase systems, which would ensure the best property performance.
To achieve this goal, we introduce a new tessellation-based procedure that ensures the creation of perfectly hyperuniform dispersions in two and three dimensions for any sample size, including the infinite-sample-size limit. 

We begin by stating preliminary definitions and computational methods (see Sec. \ref{sec:definitions}). Then, we introduce a tessellation-based procedure to construct  disordered hyperuniform \textit{dispersions}, i.e., two-phase systems in which hard (nonoverlapping) ``particles'' are spatially distributed throughout a connected ``matrix'' phase.
This procedure is a new and special type of packing protocol that is different from previously known techniques \cite{Torquato2018_packing}.
We first describe this procedure and then prove that such constructions produce perfectly hyperuniform dispersions in the infinite-sample-size limit (see Sec. \ref{sec:theory}).
As a proof-of-concept, we numerically verify our analytical results by constructing hyperuniform dispersions from two types of initial tessellations: Voronoi tessellations (see Sec. \ref{sec:byVoronoi}) and multiscale-sphere tessellations (see Sec. \ref{sec:multiscale}). 
The former case provides a remarkable mapping that converts a very large nonhyperuniform packing (at least as large as $10^7$ particles) to a hyperuniform dispersion. 
The latter case enables us to establish for the first time the hyperuniformity of the famous optimal multiscale coated-spheres model.  
The challenging numerical task of constructing such disordered multiscale dispersions is carried out and their hyperuniformity is verified. 
Finally, we demonstrate how such a capability can be combined with state-of-the-art 2D photolithographic \cite{Zhao2018} and 3D printing techniques \cite{Wong2012, Tumbleston2015, Shirazi2015} to design and fabricate very large disordered hyperuniform materials (see \ref{sec:Fabrication}).
We make concluding remarks in Sec. \ref{sec:conclusions}. 
 

\section{Preliminaries and computational methods}\label{sec:definitions}
\subsection{Spectral density} \label{sec:spectral_density} 
The microstructure of a two-phase system is often described by two-point correlation function $\fn{S_2 ^{(i)}}{\vect{r}}$ that measures the probability that two points separated by $\vect{r}$ are simultaneously located in phase $i$.
The \textit{autocovariance} function is defined as 
\begin{equation}\label{eq:autocovarianceFunction}
\fn{\chi_{_V}}{\vect{r}} \equiv \fn{S_2^{(i)}}{\vect{r}} - {\phi_i}^2,
\end{equation}
which is identical for each phase, and tends to zero as $r$ increases when the system is in the absence of long-range order \cite{Torquato_RHM}.
Its Fourier transform $\spD{\vect{k}}$ at a wavevector $\vect{k}$ is called the \textit{spectral density} and is a nonnegative real-valued function of $\vect{k}$.
Importantly, the spectral density is directly measurable from elastic scattering experiments \cite{Debye1949}.

Consider a sphere-packing in a periodic simulation box $\mathcal{V}_d$ in $d$-dimensional Euclidean space $\R^d$, which consists of $N$ spheres of different radii $R_1, R_2,\cdots,R_N$.
Its spectral density can be written as \cite{Torquato1999_3, Torquato_RHM,Torquato2016_gen}  
\begin{align}
& \spD{\vect{k}}  = \frac{1}{\abs{\mathcal{V}_d}}  \times \nonumber \\   & \abs{\sum_{j=1}^N \fn{\tilde{m}}{\abs{\vect{k}}; R_j} e^{-i\vect{k}\cdot\vect{r}_j}-\phi \int_{\mathcal{V}_d} \d{\vect{r}'} e^{-i \vect{k}\cdot\vect{r}'}}^2,\label{eq:chi_v_step1}
\end{align}
where $\phi$ is the packing fraction and $\abs{\mathcal{V}_d}$ is the volume of a simulation box.
Here, $\fn{\tilde{m}}{k;R} = \left(2\pi R/k\right)^{d/2}\fn{J_{d/2}}{kR} $ represents the Fourier transform of a sphere of radius $R$, $\fn{J_{\nu}}{x}$ is the Bessel function of order $\nu$, $L$ is side length of the unit cell, and a wavevector $\vect{k} \equiv \frac{2\pi}{L} (n_1, n_2, \cdots, n_d)$ corresponds to the reciprocal lattice vectors of the unit cell.

In numerical simulations, Eq. \eqref{eq:chi_v_step1} can be simplified as follows:
\begin{equation}
\spD{\vect{k}}  = \frac{1}{\abs{\mathcal{V}_d}} \abs{\sum_{j=1}^N \fn{\tilde{m}}{\abs{\vect{k}}; R_j} e^{-i\vect{k}\cdot\vect{r}_j}}^2,\label{eq:chi_v_simple}
\end{equation}
using the fact that the space integral in Eq. \eqref{eq:chi_v_step1} effectively vanishes at every nonzero wavevector.
We note that Eqs. \eqref{eq:chi_v_step1} and \eqref{eq:chi_v_simple} cannot be applied to overlapping spherical particles.
For an ensemble of multiple packings, the spectral density is computed by averaging spectral densities of individual packings, and the associated errors are estimated from the sample standard deviations.
The angular-averaged spectral density, of central concern in this paper, is computed by averaging the spectral densities at wavevectors whose magnitudes are in the same bin, and error bars in the wavenumbers represent the standard deviations.

\subsection{Classes of hyperuniform two-phase systems}\label{sec:HU def}
For disordered hyperuniform two-phase systems, the spectral density $\spD{\vect{k}}$ frequently exhibits power-law scaling behavior in the small wavenumber limit:
\begin{equation}\label{def:HU_spectral}
 \spD{\vect{k}} \sim \abs{\vect{k}}^\alpha~~~(\abs{\vect{k}} \to 0).
\end{equation}
Values of the positive exponent $\alpha$  define three classes of hyperuniformity associated with the large-$R$ behaviors of local volume-fraction variance \cite{Zachary2009a,Torquato2018_review,Durian2017}:
\begin{equation*}
\vv{R} \sim 
\begin{cases}
R^{-(d+1)}, & \alpha>1~(\mathrm{Class~I})\\
R^{-(d+1)}\ln{R}, & \alpha = 1~(\mathrm{Class~II})\\
R^{-(d+\alpha)}, &\alpha < 1~(\mathrm{Class~III})
\end{cases},
\end{equation*}
where class I is the strongest form of hyperuniformity.

\subsection{Random sequential addition (RSA) procedure}\label{sec:RSA}
RSA is a time-dependent process that irreversibly, randomly, and sequentially places nonoverlapping spheres into space \cite{Wisdom1966, Torquato_RHM}.
In its infinite-time limit, the resulting packing becomes saturated (i.e., there is no available space to add other particles) and yet nonhyperuniform \cite{Zhang2013}. 
We generate exactly saturated RSA packings in a finite amount of time by using the voxel-list algorithm, developed by Zhang and Torquato \cite{Zhang2013}.

\subsection{Voronoi tessellations}\label{sec:Voronoi}
For a point pattern, the Voronoi cell associated with a given point is the region of space closer to this point than to any other points, and the Voronoi tessellation is the collection of all Voronoi cells \cite{Torquato_RHM}.
While Voronoi tessellations can be naturally generalized to Laguerre \cite{Gellatly1982,Hiroshi1985} and Manhattan  \cite{Krause_TaxicabGeometry} distances, in this work we focus on Euclidean distances.

We compute the Voronoi tessellation of a given point pattern by the VORO++ library \cite{voro++}.
To improve the computational time, we divide a point pattern into several $d$-dimensional hypercubic domains with overlapping marginal regions whose thickness is $6a$, where $a\equiv \rho^{-1/d}$; see Fig.1 in Supplementary Material.
Then, Voronoi tessellations of individual domains are computed in parallel. Here, we note that in 2D implementations, the particle number in each domain should be smaller than $10^5$ to avoid any possible memory leakage, which may occur because the VORO++ library is designed for 3D Voronoi tessellations.

\subsection{Simulated sphere tessellations}\label{sec:sphere-tessellation}

A sphere tessellation is a partition of space ($\R^d$) by nonoverlapping spheres with a polydispersity in size down to the infinitesimally small.
Such multiscale-sphere tessellations are infinitely degenerate with varying degrees of order/disorder.
The most ordered ones would be derived from lattice packings or certain deterministic procedures, such as \textit{Apollonian gaskets}.
In this work, we focus on disordered sphere tessellations that are constructed by implementing a multicomponent version of the aforementioned RSA packing procedure,  which is a multi-stage process.

Specifically, this multi-stage RSA process has a control parameter and a control function, namely, an upper bound $v_{\max}$ on cell volumes and a positive decreasing function $\fn{g}{i}$ for natural numbers $i$, where $\fn{g}{1}=1$ and its infinite sum exists (i.e., $\sum_{i=1}^\infty \fn{g}{i} < \infty$).
Using these factors, we determine the prescribed number $N$ of spheres that are inserted in each stage, and the largest cell volume $v^{(1)}$ to fill all space:
\begin{align}
N & \equiv \ceil{ \abs{\mathcal{V}_d} / \left(v_{\max} \sum_{m=1}^\infty \fn{g}{m}	\right)},~\mathrm{and} \\
\abs{\mathcal{V}_d} & = N v^{(1)}\sum_{i=1}^\infty \fn{g}{i}, 
\end{align}
where $\ceil{x}$ is the ceiling function and $\abs{\mathcal{V}_d}$ is the volume of the simulation box.
In the $m$th stage, the sphere has volume $v^{(m)}= v^{(1)}\fn{g}{m}$ and diameter $D_m$, and the prescribed covering fraction is $N \sum_{i=1}^m v^{(i)} / \abs{\mathcal{V}_d}$. 

Using these parameters, sphere tessellations are constructed by the following steps:
\begin{enumerate}
\item Begin with an empty hypercubic simulation box in $\R^d$, whose volume is $\abs{\mathcal{V}_d}$, under the periodic boundary conditions.

\item In the first stage ($m=1$), $N$ nonoverlapping spheres of an identical volume $v^{(1)}$ (i.e., diameter $D_1$) are irreversibly, sequentially, and randomly added in the simulation box (i.e., RSA procedure).
The insertion in this stage stops only when the packing reaches to a prescribed covering fraction $N v^{(1)}/\abs{\mathcal{V}_d}$, unless the packing becomes saturated.
At the end of the first stage, $N_1$($\leq N$)  spheres are added and the associated covering fraction is $\eta_1\equiv N_1 v^{(1)}/\abs{\mathcal{V}_d}$.

\item In the $m$th stage ($m>1$), nonoverlapping spheres of an identical volume $v^{(m)} (< v^{(m-1)})$ are inserted via the RSA procedure into the packing generated up to the $(m-1)$st stage.  
Here, the spheres in the $m$the stage also should be nonoverlapping to those in the previous stages. 
The insertion stops when the packing reaches to a prescribed covering fraction $N\sum_{i=1}^m v^{(i)}/\abs{\mathcal{V}_d} $, unless the packing becomes saturated.
At the end of this stage, $N_m$ spheres are additionally inserted, and thus the packing has $\mathcal{N}_m (\equiv \sum_{i=1}^m N_i)$ spheres in total and a covering fraction $\eta_m = (\sum_{i=1}^m N_i v^{(i)} /\abs{\mathcal{V}_d})$.

\item This procedure is repeated until it reaches to a prescribed number of stages. 
  
\end{enumerate}
In the RSA procedure of steps 2 and 3, we implement the aforementioned voxel-list algorithm \cite{Zhang2013}.
In this work, we consider a power-law scaling  $\fn{g}{m} = 1/m^p$ whose infinite series converges to the Riemann zeta function $\fn{\zeta}{p}$, and thus schedule covering fraction in the $m$th stage is $\sum_{n=1}^m n^{-p}/\fn{\zeta}{p}$. 
We choose scaling exponents $1<p<2$ because at scaling exponent higher than 2 makes it more difficult to insert additional particles as the number $m$ of stages increases \cite{Shier2013}; see the Supplementary Material for details of the simulation parameters employed. 

\section{Theoretical Analyses}\label{sec:theory}
In this section, we precisely describe the tessellation-based procedure. 
We analytically prove that the constructed dispersions are strongly hyperuniform by deriving the small-$\abs{\vect{k}}$ scaling of the spectral densities for spherical particles and subsequently provide an intuitive rationale of the theoretical results.
  
\begin{figure}[h]
\centering
\includegraphics[width = 0.45\textwidth]{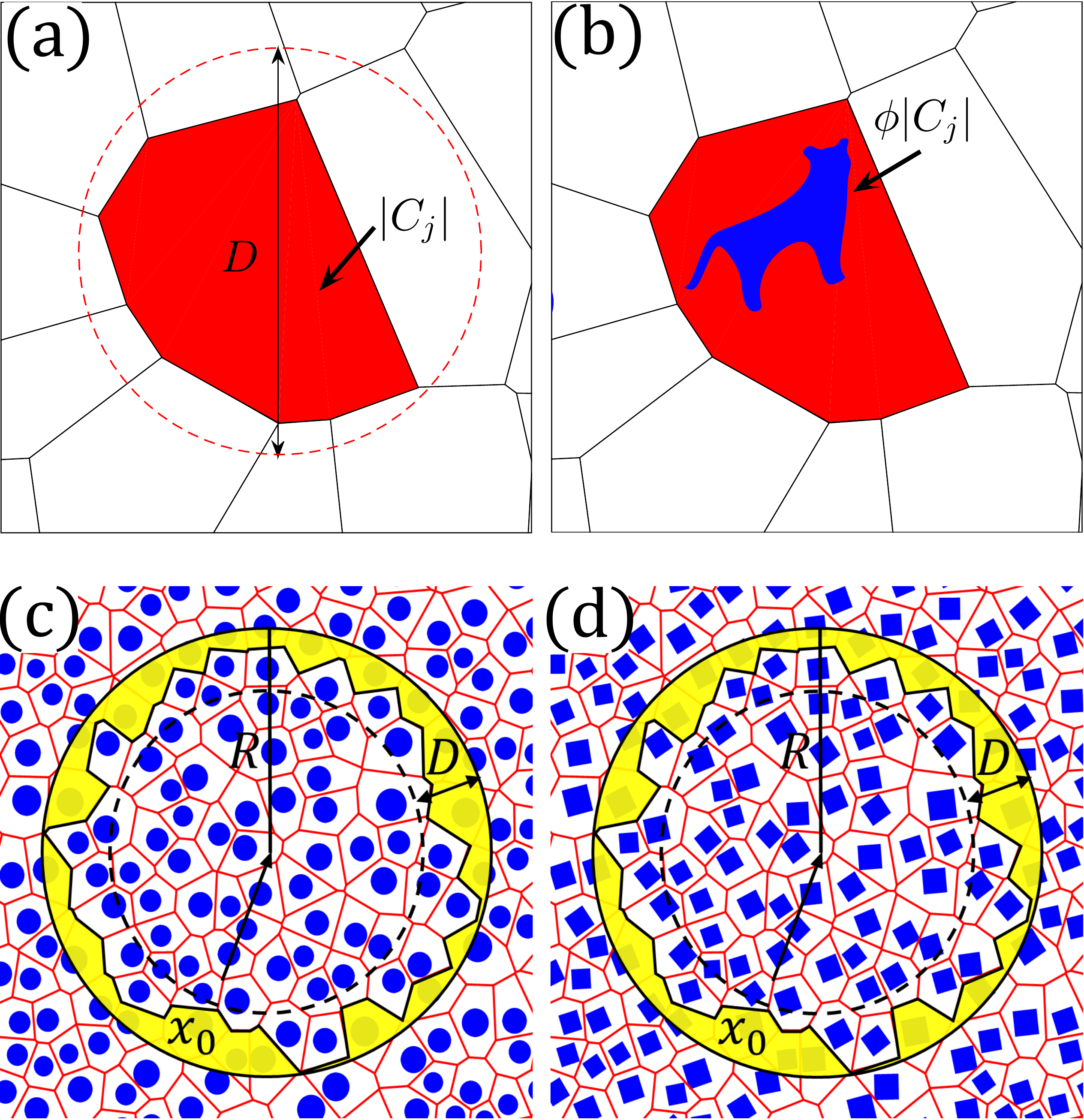}
\caption{Schematics illustrating the tessellation-based procedure to construct disordered hyperuniform dispersions. 
(a) One first tessellates the space with disjoint cells whose maximal lengths are shorter than a certain length scale $D(\ll L)$, where $\abs{C_j}$ represents the volume of cell $j$.
(b) For a specified packing fraction $0<\phi <1$, within each cell $C_j$ 
one places a single particle of general shape of volume $\phi\abs{C_j}$.
This ensures that the local-cell packing fraction $\phi$ is identical to the global packing fraction of the resulting dispersion.
(c) A special case of constructed dispersions with circular disks and a circular observation window of large radius $R(\gg D)$.
The volume-fraction fluctuations arise only inside a narrow yellow-shaded region whose effective thickness is thinner than $D$, and thus the resulting fluctuations are on the order of $R^{d-1}$, i.e., class I hyperuniformity.
(d) The corresponding hyperuniform dispersion of randomly oriented squares.
The same rationale explains the hyperuniformity of the dispersion of squares (d) as well as the Hashin-Shtrikman multiscale coated-disks structures shown in Fig. \ref{fig:multiscale_tiling_initial}(b).
\label{fig:schematics}
}
\end{figure} 

\subsection{Tessellation-based procedure}\label{sec:procedure}
Consider a hypercubic simulation box of side length $L$ in $d$-dimensional Euclidean space $\R^d$ under the periodic boundary conditions.
Our procedure, graphically illustrated in Fig. \ref{fig:schematics}, consists of the following steps:
\begin{enumerate}
\item  Divide the simulation box into $N$ disjoint cells whose maximal lengths must be smaller than a given length scale $D$, which also should be much smaller than $L$; see Fig. \ref{fig:schematics}(a).
\label{step1}
\item
Fill the $j$th cell $C_j$ with a single particle or multiple particles of general shape of total volume $\phi \abs{C_j}$; see Fig. \ref{fig:schematics}(b) for the simple instance of a single  particle per cell.  
Then, repeat the same process over all cells.
The global packing fraction of the constructed dispersion is identical to the local-cell packing fraction $\phi$. 
 \label{step2}
\end{enumerate}
We call the restriction on the cell size in step \ref{step1} the \textit{bounded-cell condition}. 
We note that given an initial tessellation, our construction is realizable only when the local-cell packing fraction $\phi$ in step \ref{step2} is smaller than or equal to the maximal packing fraction $\phi_{\max}$, i.e.,
\begin{equation}\label{eq:max_vol_fraction}
\phi \leq \phi_{\max}\equiv \min_{j=1}^N \left\{ \frac{\fn{v_1}{R^\mathrm{max}_j}}{\abs{ C_j}  }	\right\},
\end{equation}
where $\abs{ C_j}$ and $R^\mathrm{max}_j$ represent volume of the $j$th cell and radius of the largest sphere inscribed  in this cell, respectively.
Roughly speaking, the maximal packing fraction $\phi_{\max}$ becomes larger when the cell shapes become more isotropic (sphere-like).

\subsection{General theoretical analyses} \label{sec:theoretical analyses}

Here, we sketch the proof that the constructed dispersions are hyperuniform, the details of which are provided in Ref. \cite{Jaeuk2018_4}. 
We start from a general expression \eqref{eq:chi_v_step1} for the spectral density of a sphere packing of various particle radii \cite{Torquato1999_3, Torquato_RHM,Torquato2016_gen}. 
We now decompose $\phi \int_{_{\mathcal{V}_d}} \d{\vect{r}'} e^{-i \vect{k}\cdot\vect{r}'}$ into the Fourier transform of each cell, and rearrange each term with that of the associated particle.
Importantly, the bounded-cell condition guarantees that the Fourier transforms of both a cell $C_j$ and the associated particle $j$ can be well approximated by their Taylor series about $\vect{k}=\vect{0}$. 
It follows that the tessellation-based procedure ensures that the leading order terms in these series exactly cancel one another such that the remaining terms exhibit power-law scalings in the wavenumber, i.e., $\abs{\vect{k}}^\alpha$ as $\abs{\vect{k}}\to 0$.  
A detailed analysis given in Ref. \cite{Jaeuk2018_4}  
shows that in the small-$\abs{\vect{k}}$ limit, $\spD{\vect{k}} \sim \phi^2\abs{\vect{k}}^2$ is achievable, in general.
However, the behavior $\spD{\vect{k}}\sim \phi^2\abs{\vect{k}}^4$ is achieved when the special condition $\vect{k}\cdot \left[	\sum_{j=1}^N \Delta \vect{r}_j \fn{v_1}{R_j} e^{-i\vect{k}\cdot\vect{r}_j} \right] = 0$ is satisfied, where $\Delta \vect{r}_j$ represents the displacement between the centroids of cell $j$ and the associated particle.
These results clearly show that the constructed dispersions are hyperuniform of class I for $\phi \leq \phi_{\max}$, given in Eq. \eqref{eq:max_vol_fraction}, which also guarantees that the particles remain inside the associated cells.
Note that $\spD{\vect{k}}$ is exactly the same for either the particle or matrix (space exterior to particles) phases \cite{Torquato_RHM} and hence the matrix phase is also strongly hyperuniform.

To understand intuitively why a dispersion constructed by this procedure is hyperuniform, it is useful to consider the local volume-fraction variance $\vv{R}$.
Imagine sampling the constructed dispersion with many randomly placed observation windows of radius $R(\gg D)$.
Clearly, the volume-fraction fluctuations will be concentrated only in the small region near the boundary of the window, as shown in Fig. \ref{fig:schematics}(c).
Consequently, the variance in the phase-volume will be proportional to the window surface area in the large-$R$ limit, i.e., $ {\fn{v_1}{R}}^2 \vv{R} \sim R^{d-1}$, and in turn $\vv{R}\sim R^{-(d+1)}$, implying that such dispersions are strongly hyperuniform (class I). Our methodology is valid for any particle shape as shown in Fig. \ref{fig:schematics}(d) and allows for the addition of multiple particles per cell provided that local-cell packing fraction is identical for each cell. However, for concreteness, we mainly analyze dispersions that have only a single spherical particle in each cell.

\section{Disordered hyperuniform dispersions derived from Voronoi tessellations}\label{sec:byVoronoi}
\begin{figure}[h]
\centering
\includegraphics[width = 0.45\textwidth]{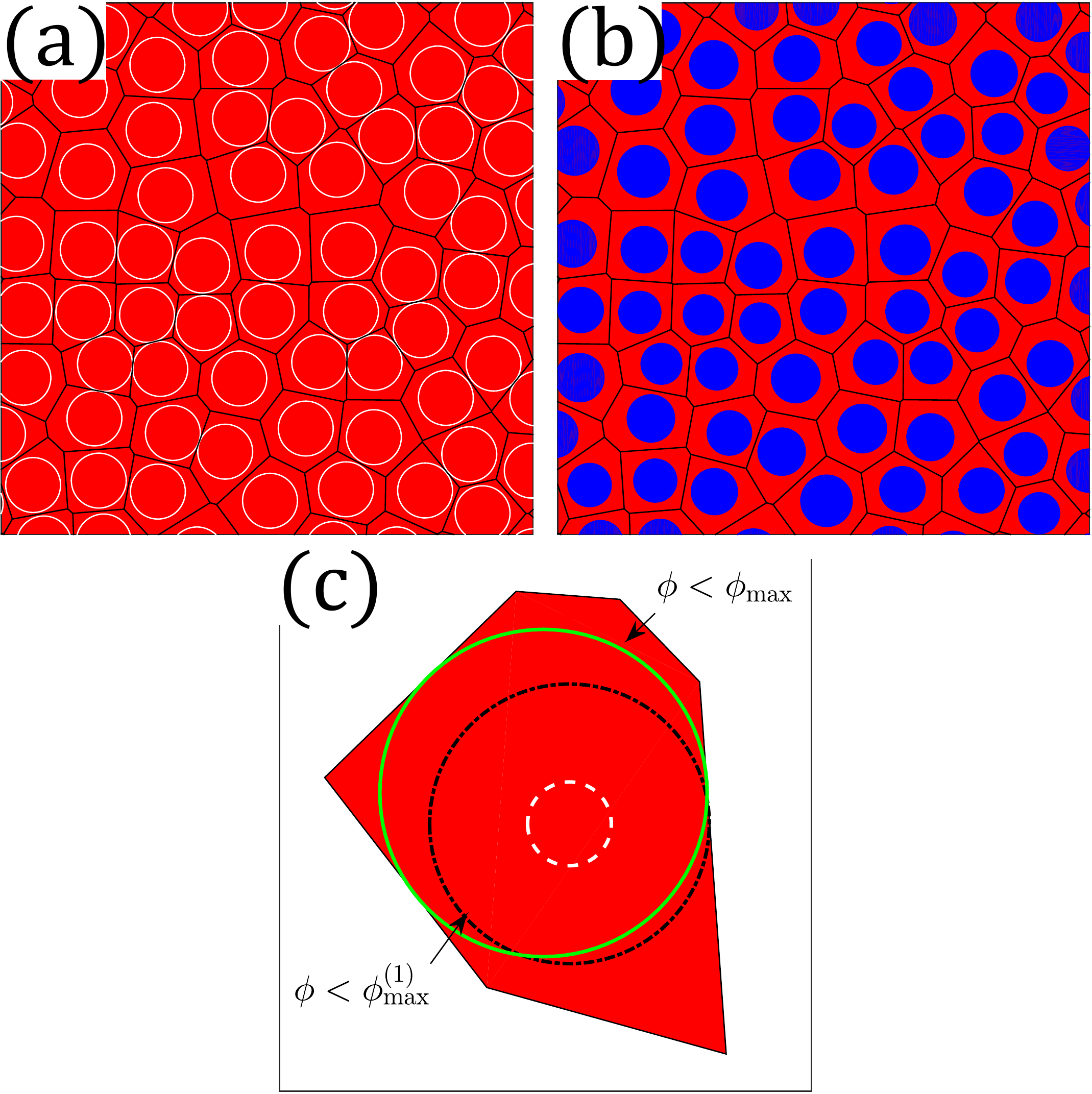}
\caption{Implementation of the tessellation-based methodology via Voronoi tessellations.
(a) A portion of Voronoi tessellation of a progenitor packing (i.e., a 2D saturated RSA packing) of white circles.
(b) A hyperuniform dispersion (blue disks) of packing fraction $\phi$ that is constructed by our methodology from (a) with the particle centers fixed. 
(c) Schematic of two types of maximal packing fractions $\phi_{\max}^{(1)}$ and $\phi_{\max}$. 
In this cell, while at $\phi = \phi_{\max}$ the particle can be as large as the largest inscribed circle (green circle), at $\phi=\phi_{\max}^{(1)}$ the particle size can be as large as the largest inscribed circle (black) that is concentric to the initial particle (white circle). 
\label{fig:TilingProcedure2DsaturatedRSA1}}
\end{figure}

\begin{figure*}[h!t]
\includegraphics[width = 0.32\textwidth]{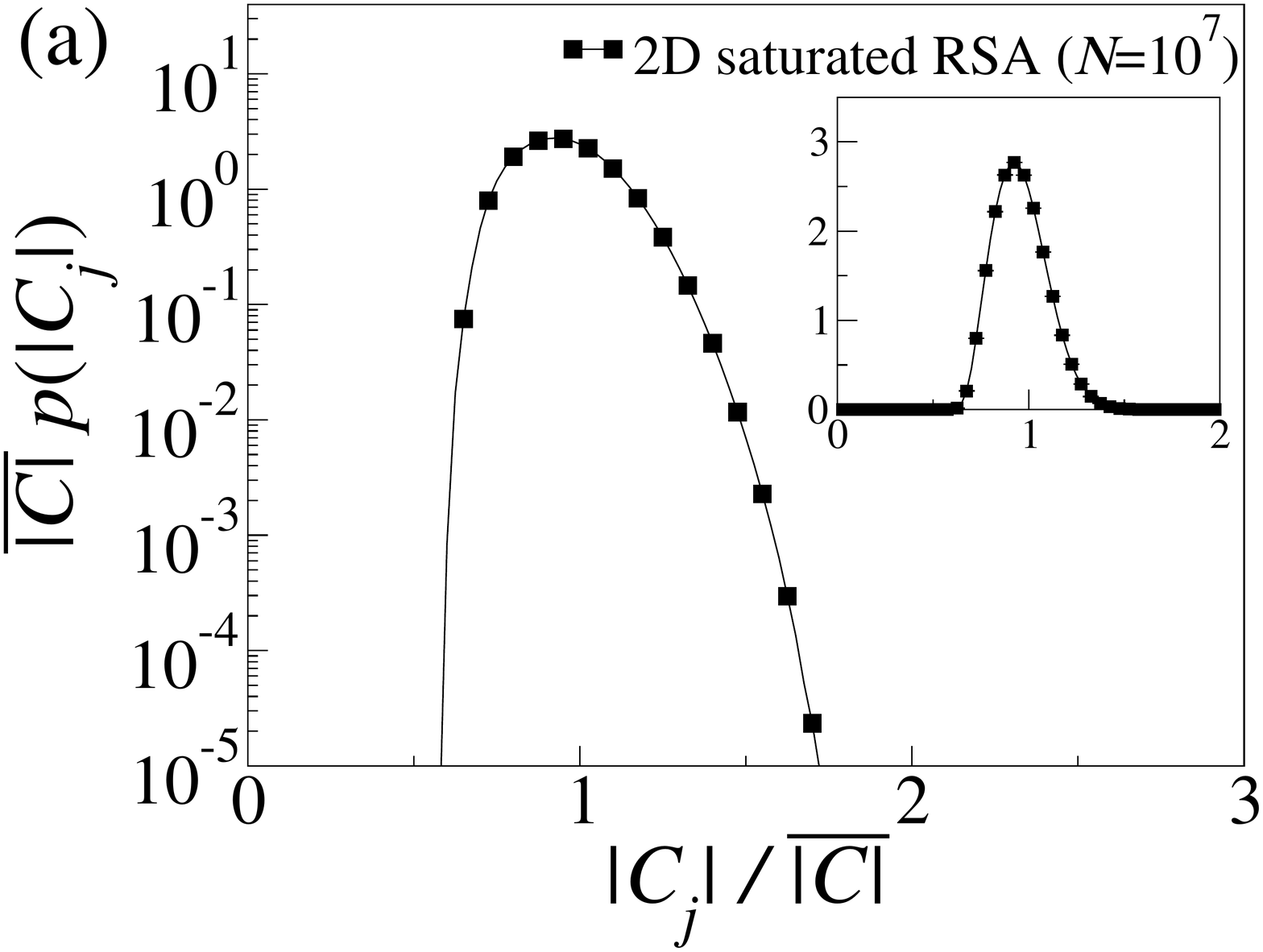}
\label{fig:voronoicellDist}
\includegraphics[width = 0.32\textwidth]{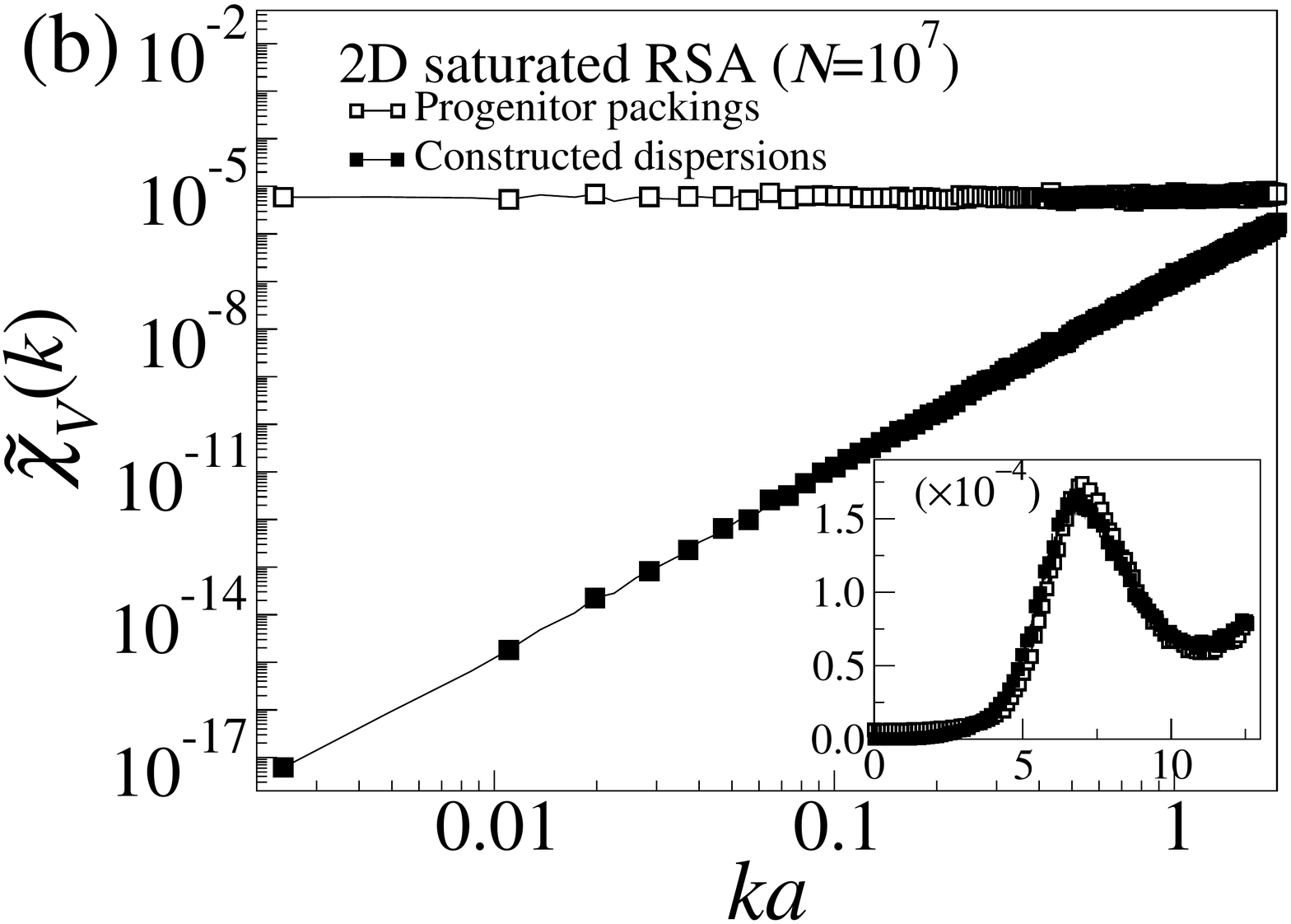}
\label{fig:2D_RSA_chivk}
\includegraphics[width = 0.32\textwidth]{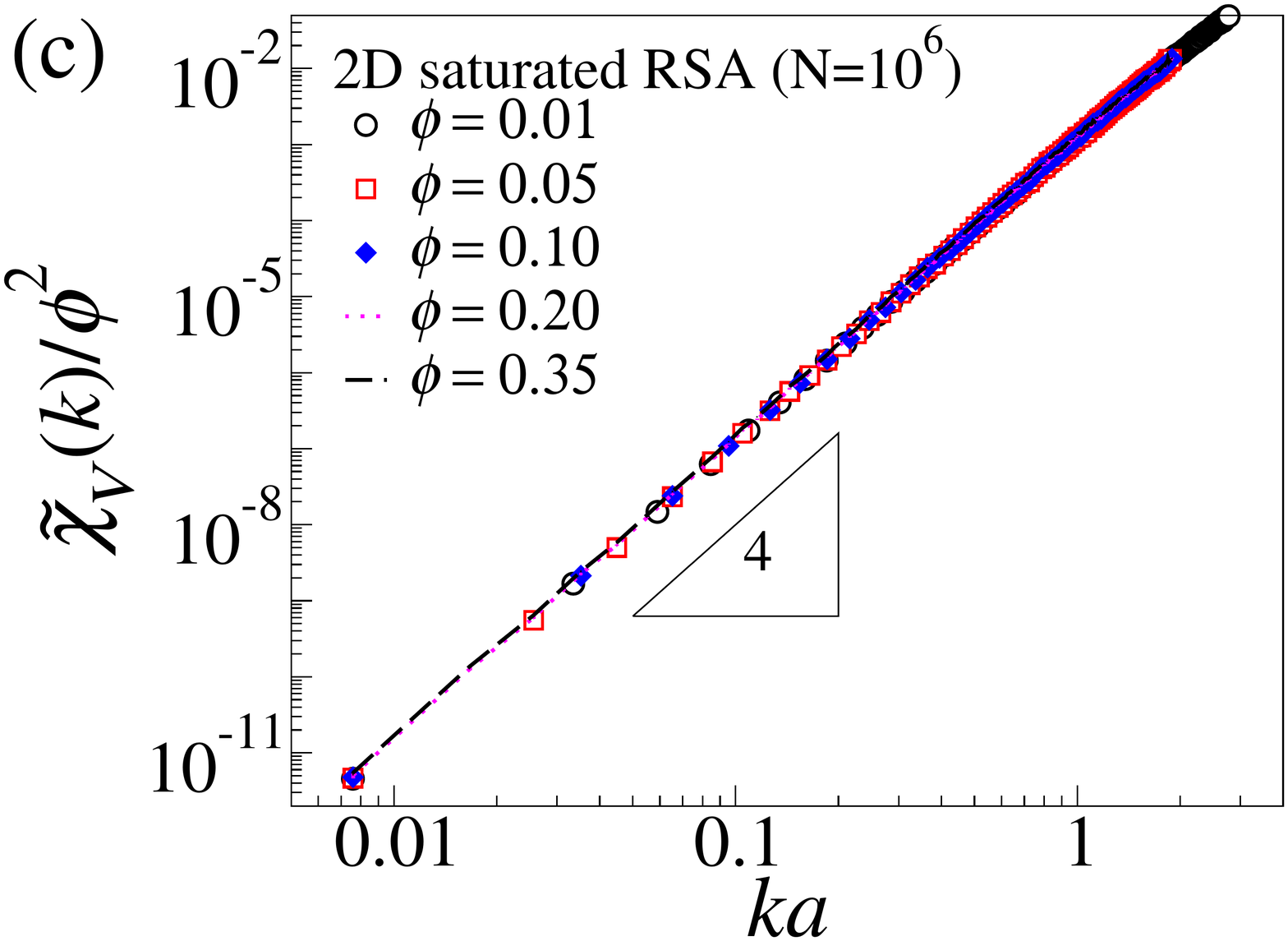}
\label{fig:2D_RSA_phi_dependence}

\includegraphics[width = 0.32\textwidth]{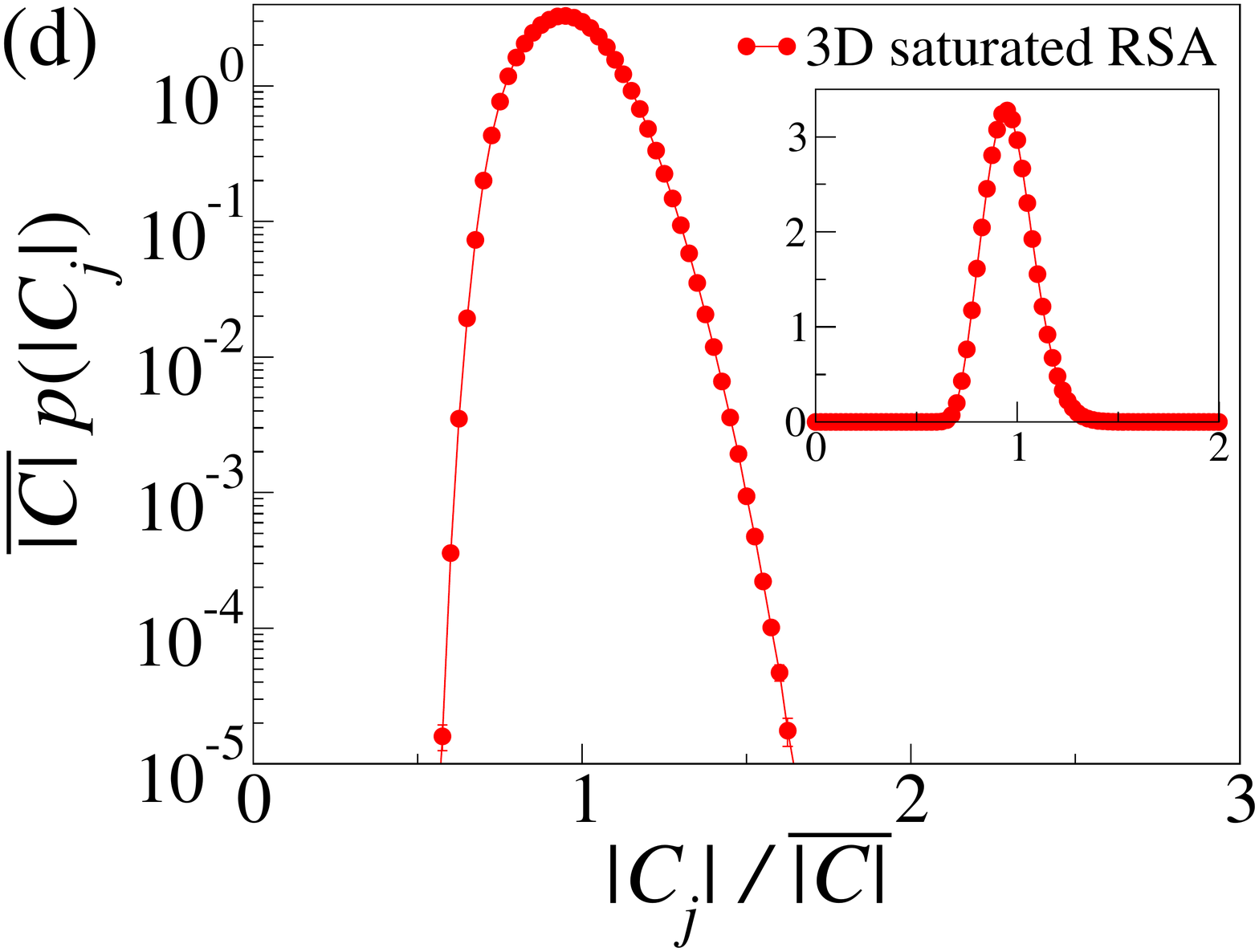}
\label{fig:3DvoronoicellDist}
\includegraphics[width = 0.32\textwidth]{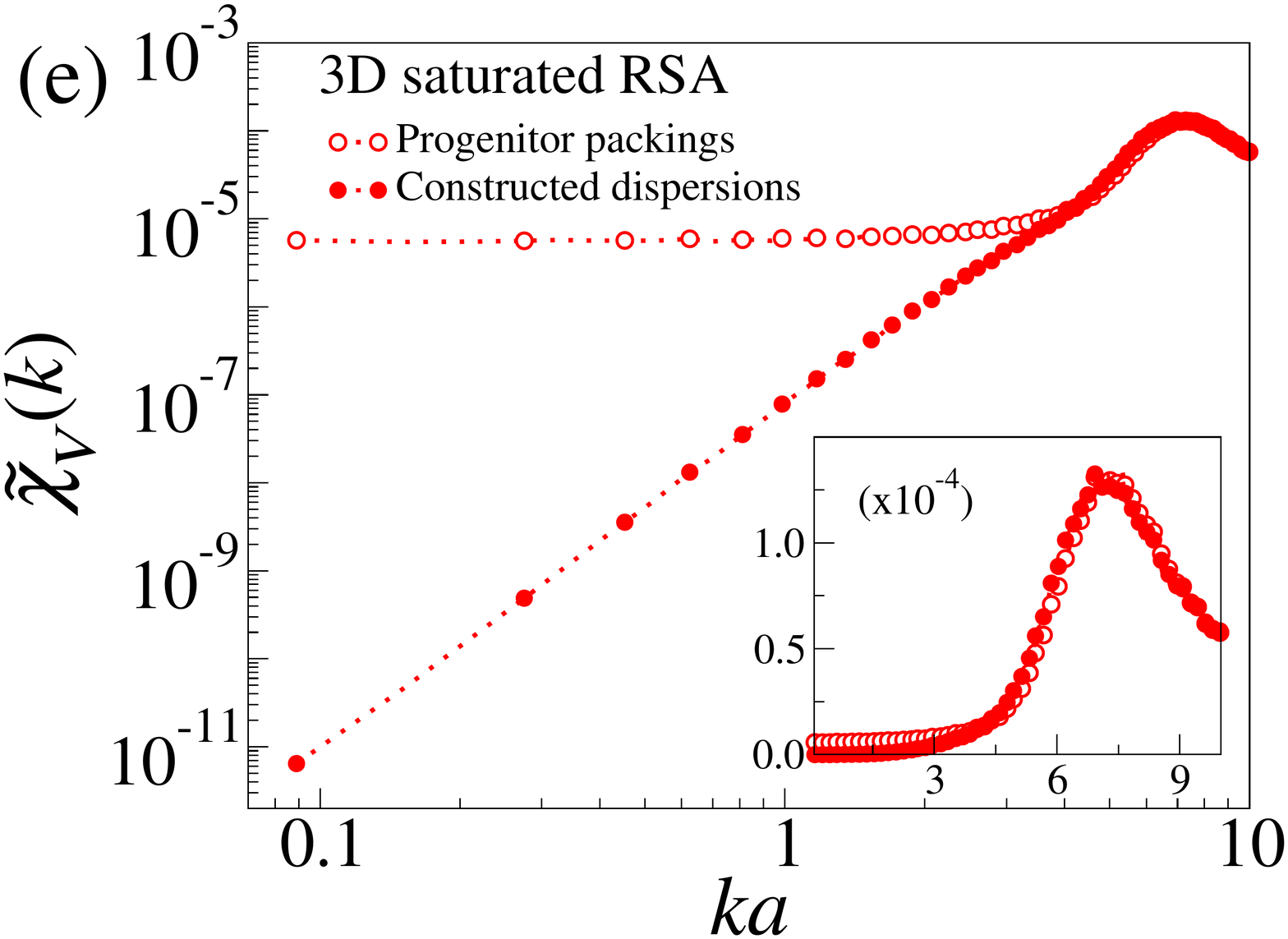}
\label{fig:3D_RSA_chivk}
\includegraphics[width = 0.32\textwidth]{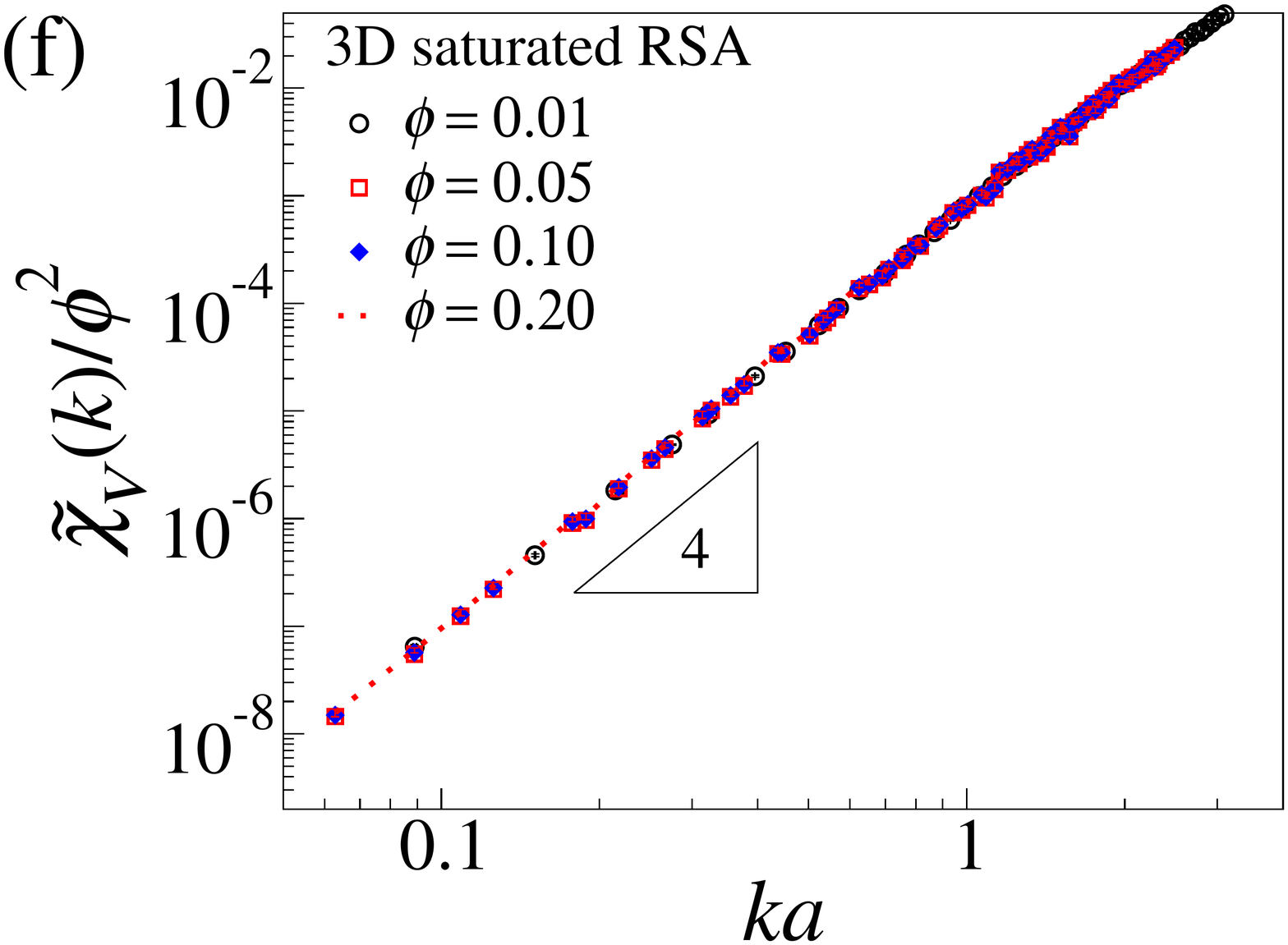}
\label{fig:3D_RSA_phi_dependence}

\caption{Simulation results of dispersions constructed from saturated RSA packings in $\R^2$ and $\R^3$.
(a) The probability density function of Voronoi cell volume $\abs{C_j}$ of 2D saturated RSA packings on a semi-log scale (larger panel) and a linear scale (inset).
Here, $\overline{\abs{C}}$ represents the average cell volume.
Error bars represent standard deviations (see Sec. \ref{sec:spectral_density}).  
(b) Spectral densities of 2D saturated RSA packings and the constructed dispersions on a log-log (larger panel) and a linear scale (inset), where $a\equiv\rho^{-1/d}$, $d$ is space dimension, and $\rho$ is the number density of particle centers.
We note that both progenitor packings and the constructed dispersions are scaled to the packing fraction $\phi=0.01$.
(c) Log-log plot of normalized spectral densities $\spD{k}/\phi^2$ of the constructed dispersions for various local-cell packing fractions $\phi$.
(d-f) Corresponding results of constructed dispersions from 3D saturated RSA packings. 
\label{fig:TilingProcedure2DsaturatedRSA2}}
\end{figure*}

As a proof-of-concept, we now implement the tessellation-based procedure numerically via Voronoi tessellations (see Sec. \ref{sec:Voronoi}) of disordered and ``nonhyperuniform'' sphere packings; see Fig. \ref{fig:TilingProcedure2DsaturatedRSA1}.
Thus, performing our methodology from Voronoi tessellations provides an efficient mapping that remarkably converts very large ($N\sim 10^7$) nonhyperuniform progenitor point patterns to perfectly hyperuniform dispersions.

To properly implement our methodology, the Voronoi tessellations of the progenitor point patterns should obey the bounded-cell condition of step \ref{step1}.
Surprisingly, when sample sizes are sufficiently large, this condition is satisfied by Voronoi tessellations of virtually all statistically homogeneous point patterns, including Poisson point patterns.
This seems counterintuitive because in the infinite-sample-size limit Poisson point patterns as well as many other disordered ones can possess arbitrarily large holes, or equivalently, arbitrarily large Voronoi cells.
Nonetheless, a detailed analysis given in Ref. \cite{Jaeuk2018_4} 
definitely demonstrates that practically all samples of these systems meet the bounded-cell condition as the sample size grows.
To sketch the basic idea, we note that for Poisson point patterns of $500,000$ points, the probability that a single finite-size sample possesses a hole larger than $0.1\%$ of its sample size is around $10^{-22}$, and this probability becomes exponentially lower as sample size grows.
Interestingly, while such rare events of large hole formation play a crucial role in many physical phenomena, such as the diffusion-controlled reactions \cite{Torquato1991},  the bounded-cell condition is not governed by these rare events.
This has the crucially practical implication that any statistically homogeneous point pattern, whether hyperuniform or not, can be used to numerically implement our mapping, which is consistent with a recent study of random fields generated from Voronoi tessellations for the special case of a Poisson point process \cite{Kampf2015}.

In this paper, we choose to perform the tessellation-based procedure from Voronoi tessellations of saturated RSA packings; see Fig. \ref{fig:TilingProcedure2DsaturatedRSA1}(a) and Sec. \ref{sec:Voronoi}. 
In contrast to uncorrelated Poisson point patterns, saturated RSA packings unquestionably meet the bounded-cell condition for even relatively \textit{small} sample sizes (say, $10^{d+1}$ particles in $d$ dimensions).   
From saturated RSA packings via the voxel-list algorithm \cite{Zhang2013},  for simplicity we construct dispersions (Fig. \ref{fig:TilingProcedure2DsaturatedRSA1}(b)) by solely scaling particle sizes without changing their positions, and hence their maximal packing fractions are defined by $\phi_{\max}^{(1)}$, illustrated in Fig. \ref{fig:TilingProcedure2DsaturatedRSA1}(c).

Figure \ref{fig:TilingProcedure2DsaturatedRSA2} summarizes simulation results of our methodology from Voronoi tessellations of saturated RSA packings.
We have constructed large saturated RSA packings with $N\approx 10^7$ and $10^6$ particles in $d=2$ and $3$, respectively.
For these progenitor packings, the maximal packing fractions $\phi_{\max}^{(1)}$ are computed from Eq. \eqref{eq:max_vol_fraction} by replacing $R_j ^{\max}$ with a half of the nearest-neighbor distance of particle $j$.
The corresponding values of $\phi_{\max}^{(1)}$ are $0.360(4)$ and $0.252(7)$ in $d=2$ and $d=3$, respectively. 
Details of the simulation parameters employed, computational times, and machine type are summarized in the Supplementary Material.
By construction, our dispersions always have a particle-size distribution that is identical to the cell-size distribution (Figs. \ref{fig:TilingProcedure2DsaturatedRSA2}(a) and \ref{fig:TilingProcedure2DsaturatedRSA2}(d)), which can be well approximated by the Gamma or the log-normal distributions. 
Thus, from the definition of a saturated RSA packing, we can estimate bounds on the particle volumes at the local-cell packing fraction $\phi$ as follows: $ \phi_\mathrm{sat} < \frac{\fn{v_1}{R_j}}{\phi \overline{\abs{C}}}<2^d \phi_\mathrm{sat}$, where $\phi_\mathrm{sat}$ represents the saturated packing fraction of RSA in $\R^d$ and $\overline{\abs{C}}$ is the mean cell-volume. 
In fact, the largest particle volumes are about three times as large as the smallest ones in both $d=2$ and $3$.  
Finally, the spectral density exhibits the behavior $\spD{\abs{\vect{k}}}\sim \phi^2\abs{\vect{k}}^4$ as $\abs{\vect{k}}$ goes to zero (Figs. \ref{fig:TilingProcedure2DsaturatedRSA2}(b)-\ref{fig:TilingProcedure2DsaturatedRSA2}(c) and \ref{fig:TilingProcedure2DsaturatedRSA2}(e)-\ref{fig:TilingProcedure2DsaturatedRSA2}(f)). 
Thus, the constructed dispersions in both $\R^2$ and $\R^3$ are strongly hyperuniform (class I) and their deviation term $\sum_{j=1}^N \Delta \vect{r}_j \fn{v_1}{R_j}$ effectively vanishes.
Finally, it is noteworthy that the spectral densities of both progenitor and constructed dispersions collapse onto a single curve for intermediate wavenumbers; see insets in Fig. \ref{fig:TilingProcedure2DsaturatedRSA2}(b,e).
This behavior is expected, since both dispersions have identical local statistics (see Fig. \ref{fig:TilingProcedure2DsaturatedRSA2}(a) and (d)). 

\begin{figure}[th]
\centering
\includegraphics[width = 0.3\textwidth]{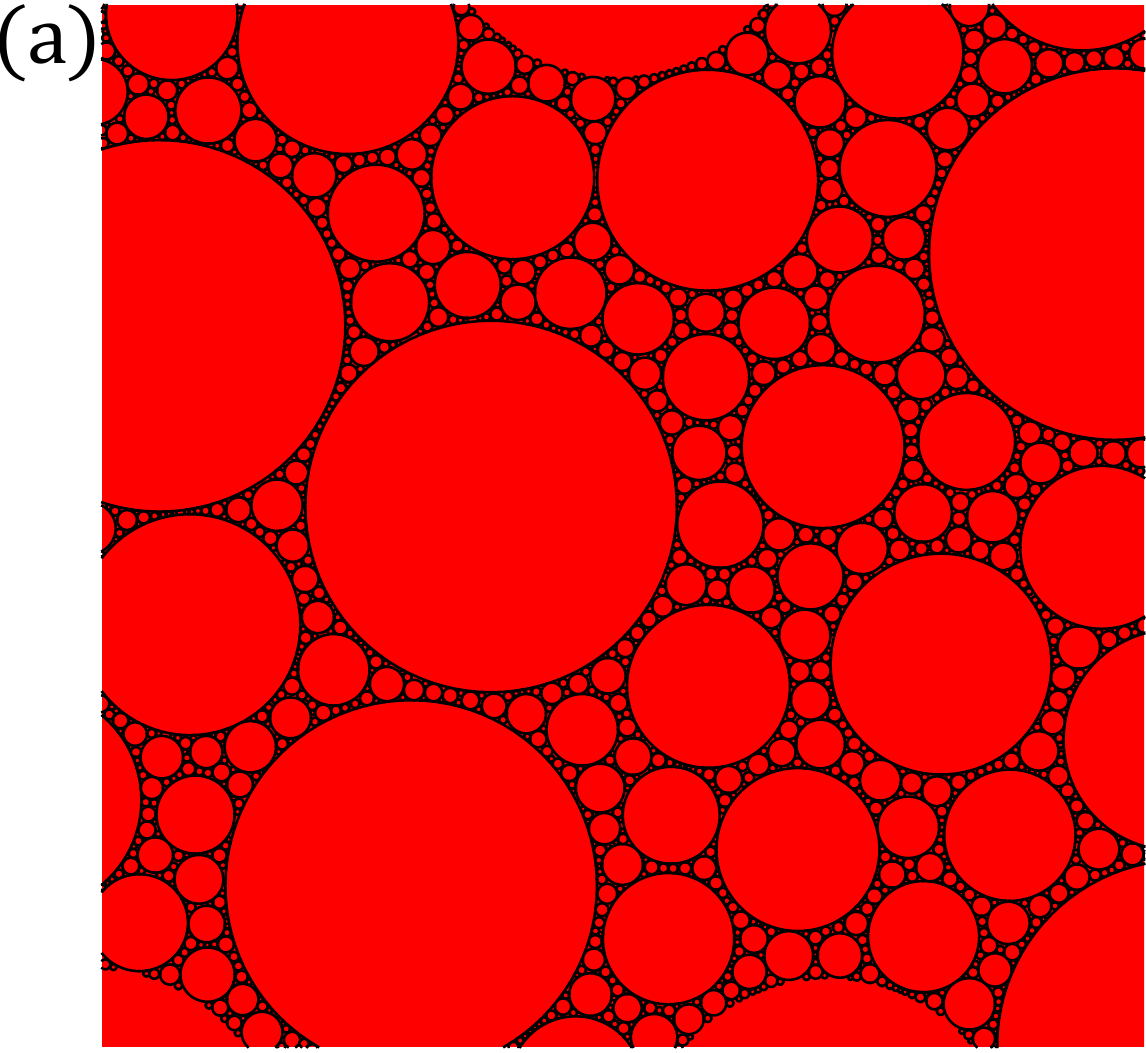} \label{fig:multiscale_tiling_initial}
\includegraphics[width = 0.3\textwidth]{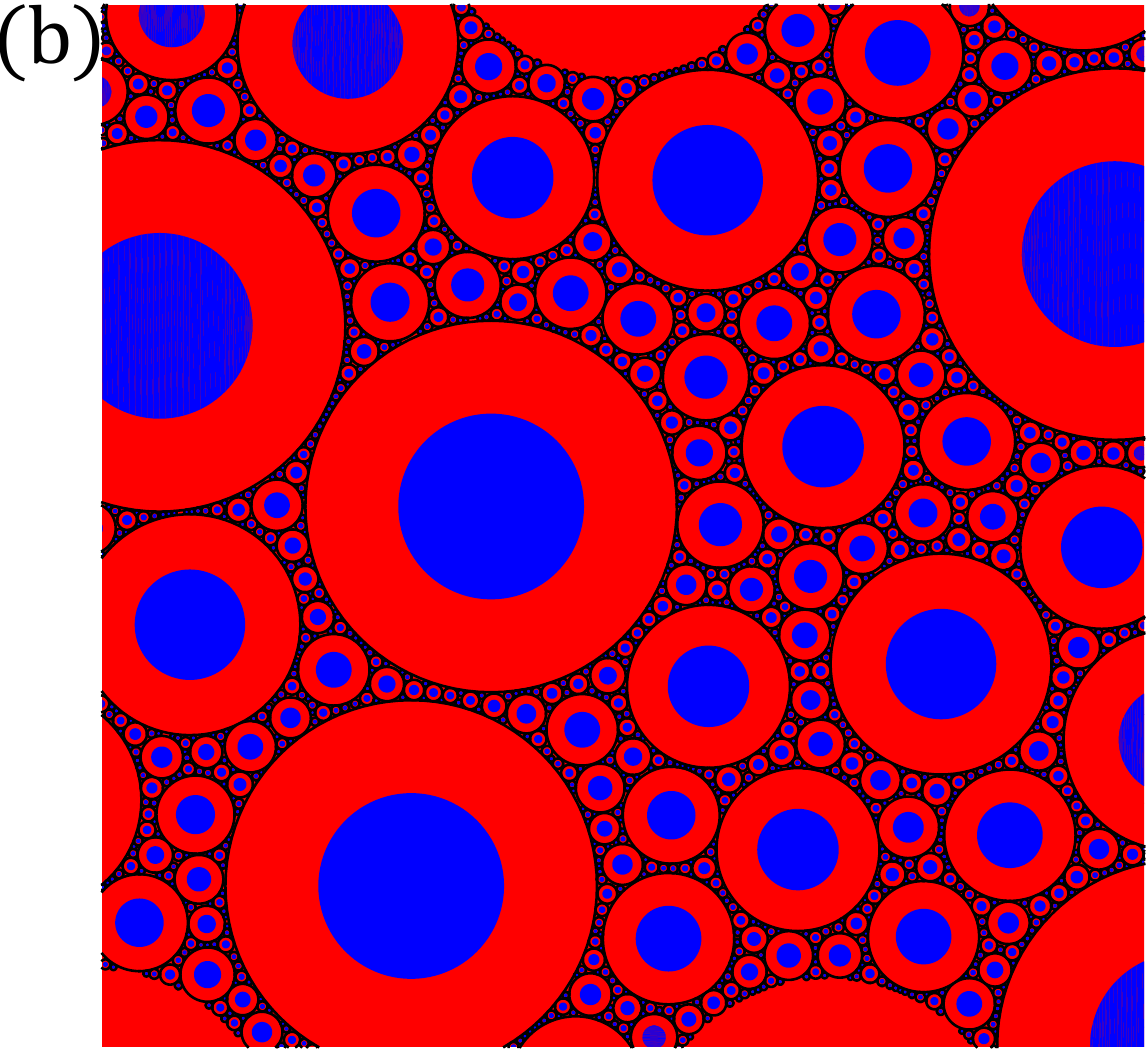}\label{fig:multiscale_tiling_final}
\caption{Illustration of the construction of  the coated-spheres model via sphere-tessellations in two dimensions.
(a) An initial multiscale-disk tiling (400th stage) and (b) a dispersion constructed via the tessellation-based procedure from (a), which is the coated-disks model. 
}
\end{figure}

\section{Hyperuniformity of the optimal coated-sphere structures}\label{sec:multiscale}
\begin{figure*}[h]
\begin{center}
\includegraphics[width =0.3 \textwidth]{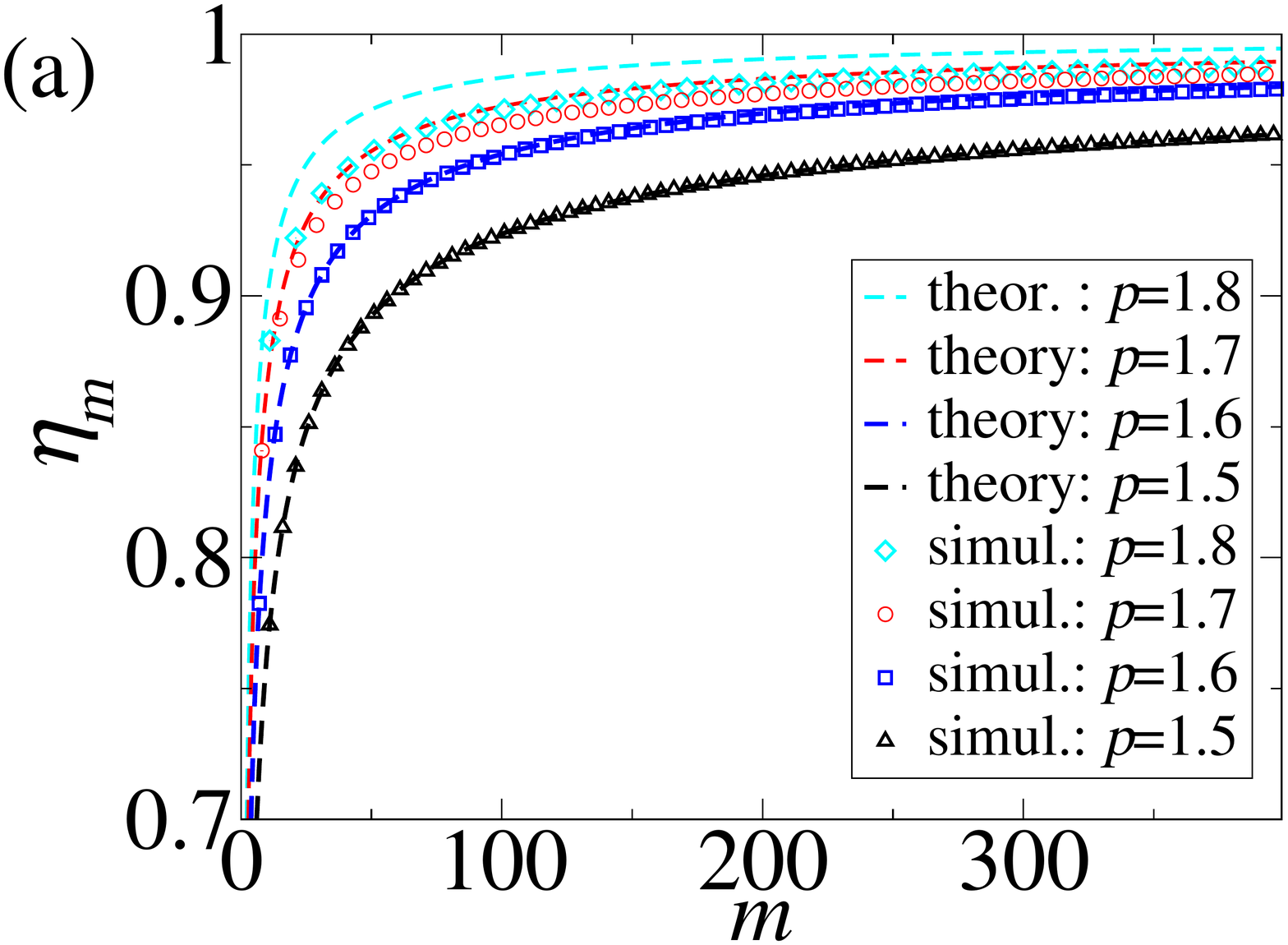}\label{fig:covering_fraction}
\includegraphics[width =0.3 \textwidth]{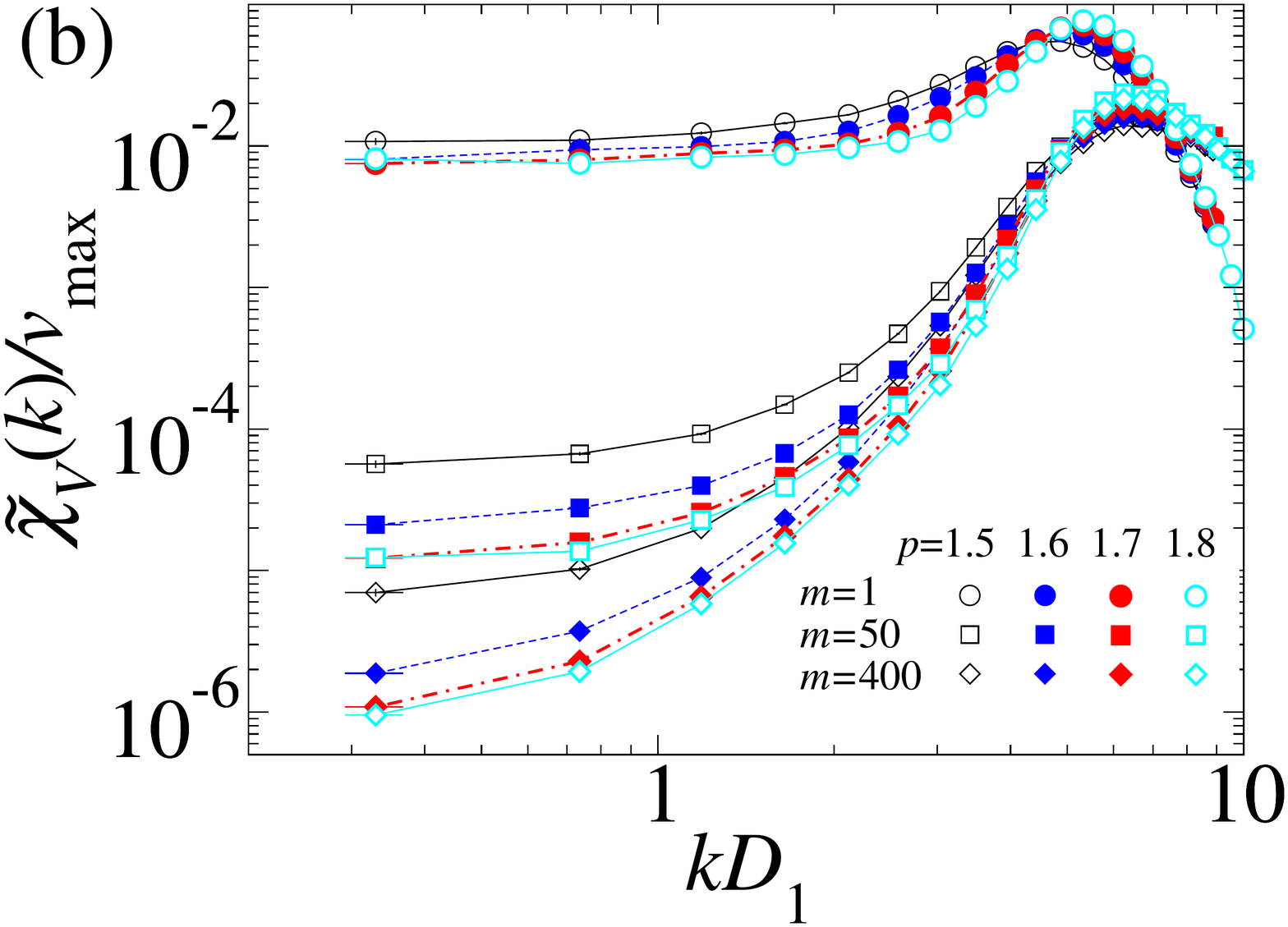}\label{fig:spectral_densities_HS}
\includegraphics[width =0.3 \textwidth]{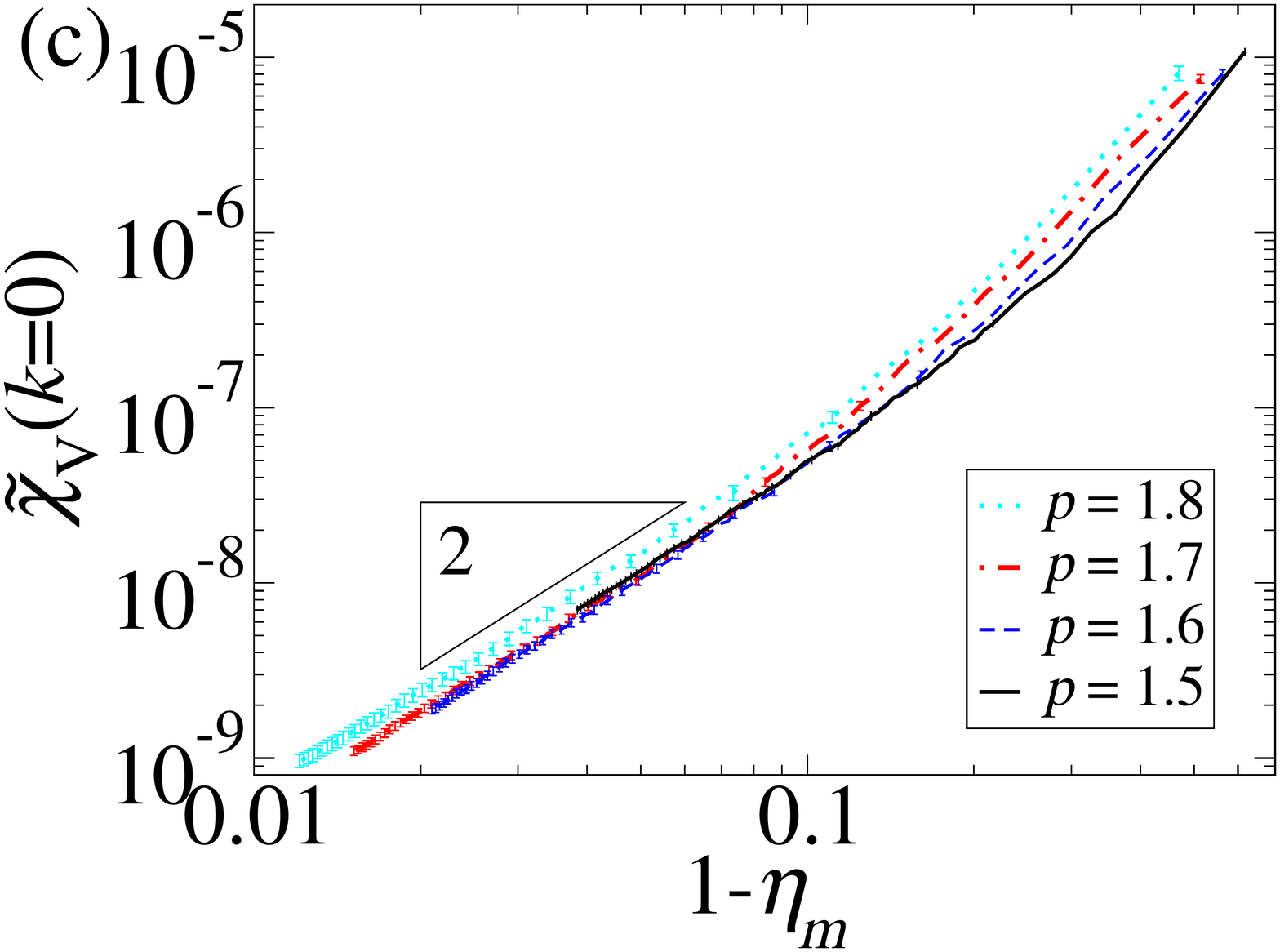}\label{fig:chiv0VScovering fraction}
\end{center}
\caption{Structural characteristics of the simulated coated-disks models (Sec. \ref{sec:sphere-tessellation}). 
(a) Covering fraction $\eta_m$ of simulated initial tilings as functions of stage number $m$. 
(b) Spectral densities for the simulated multiscale coated-disks model at various $m$'s at the local-cell packing fraction $\phi=0.5$. 
Error bars represent standard deviations (see Sec. \ref{sec:spectral_density}). (c) Spectral density at the minimal wavenumber $\spD{k\to 0}$ as a function of fraction of uncovered space $1-\eta_m$.
\label{fig:multiscale-tessellations}}
\end{figure*}

It is noteworthy that our tessellation-based methodology can be applied to other tessellations besides Voronoi tessellations that consist of polyhedral cells.
For example, although space cannot be tessellated by identical nonoverlapping spheres, it can be tessellated by nonoverlapping spheres with a polydispersity in size down to the infinitesimally small; see, for example, the multiscale-sphere tessellation in Fig. \ref{fig:multiscale-tessellations}(a) and Sec. \ref{sec:sphere-tessellation}. 
Application of our tessellation-based procedure to multiscale-sphere tessellations should produce hyperuniform dispersions, regardless of the particle shapes and cell positions with varying degrees of order/disorder. 
We now show that the famous Hashin-Shtrikman multiscale coated-spheres structures from the theory of heterogeneous materials are a special case derived from such sphere tessellations. 
This broad class of particulate composites are optimal for the effective thermal (electrical) conductivity and elastic moduli for prescribed phase properties and volume fractions \cite{Hashin1962, Torquato_RHM, Torquato2004}.
Such an optimal two-phase material consists of composite spheres that are composed of a spherical core of one phase (dispersed phase) that is surrounded by a concentric spherical shell of the other phase such that the fraction of space occupied by the core phase is equal to its overall phase volume fraction. 
The composite spheres fill all space, implying that their sizes range down to the infinitesimally small. 
Thus, this structure is a dispersion, i.e., the matrix (red region) is a fully connected (continuous) phase and the inclusions (blue regions) are ``well-separated'' from one another \cite{Torquato2018_3, Torquato2018_5}; see Fig. \ref{fig:multiscale-tessellations}(b). 
Here, we, for the first time, prove that  these Hashin-Shtrikman multiscale optimal dispersions are in fact hyperuniform, which apparently is related to their optimal transport and elastic properties.

It is crucial to observe that in the coated-spheres model, the composite spheres comprise a multiscale-sphere tessellation (see Fig. \ref{fig:multiscale-tessellations}(a)) and the fraction of space occupied by the core phase is equal to its global phase volume fraction. 
In other words, these optimal structures can be constructed via our methodology from multiscale-sphere tessellations. 
Assuming the bounded-cell condition, this implies that the coated-spheres model should be hyperuniform, whether disordered or not.   
Further analysis shows that for the coated-spheres model of packing fraction $\phi$ in $\R^d$, the small-$\abs{\vect{k}}$ scaling of the associated spectral density will exhibit the following behavior: $\spD{\abs{\vect{k}}}\sim \left(\phi  (1-\phi^{2/d})	\right)^2 \abs{\vect{k}}^4$, implying class I; see Ref. \cite{Jaeuk2018_4} 
for details.

We carry out the challenging numerical task of constructing such disordered multiscale dispersions and verify their hyperuniformity. 
To do so, first we implement a multicomponent version of the aforementioned RSA packing procedure, which amounts to a multi-stage process; see Sec. \ref{sec:sphere-tessellation}.
Then, we simulate the coated-spheres model of the inclusion volume fraction $\phi$ by reducing sphere volumes in a precursor packing at a volume ratio of $\phi$ without moving their centers.
In a finite $m$th stage, the precursor packing will not cover all space and hence there will be gaps in which smaller spheres can be added in next stages.  
Thus, as the number of stages increases in the limit of $m\to\infty$, those gaps are eventually covered by spheres of size down to the infinitesimally small, i.e., $\lim_{m\to \infty} \eta_m =1$.
To achieve a nearly complete tiling of space (say, $\eta_m \approx 0.95$) in $d=2$, this process requires around a few hundred stages; see Fig. \ref{fig:multiscale-tessellations}(c).

To estimate the degree of hyperuniformity in the $m$th stage, we employ an upper bound of the spectral density $\fn{\tilde{\chi}_{_V}^{(m)}}{\vect{k}}$ with any prior knowledge of cell-volume distributions \cite{Jaeuk2018_4}: 
\begin{align}
\fn{\tilde{\chi}_{_V} ^{(m)}}{\vect{k}} \leq & \left(\frac{\phi(1-\phi^{2/d})}{2(2+d)}	\right)^2 \frac{2}{\abs{ \mathcal{V}_d}}\abs{\sum_{j=1}^{\mathcal{N}_m}  \fn{v_1}{\mathcal{R}_j} (k\mathcal{R}_j)^2 e^{-i\vect{k}\cdot\vect{r}_j}}^2 \nonumber \\
& + 2 \abs{\mathcal{V}_d } \phi^2 (1-\eta_m)^2,\label{eq:chi_v_coated M}
\end{align}
where $\phi$ stands for the local-cell packing fraction, and $\mathcal{R}_j$ denotes radius of cell $j$.
In short, this upper bound is obtained by assuming that complex Fourier components of all uncovered gaps constructively interfere, and thus the second term in Eq. \eqref{eq:chi_v_coated M} represents the largest possible volume-fraction fluctuations due to these gaps.  
This rigorous upper bound \eqref{eq:chi_v_coated M} indicates that in the limit of $\eta_m \to 1$, the spectral density is at least on the order of $k^4$ for small $\vect{k}$'s, implying that our coated-spheres construction becomes strongly hyperuniform in the limit of $m\to \infty$.

In what follows, we focus on numerical results in two-dimensional cases for simplicity.
The simulations proceed up to the 400th stage with the following parameters. Particle volumes in the $m$th stage is determined by a power-law scaling $\fn{v_1}{D_m/2} = \fn{v_1}{D_1/2} m^{-p}$ for a scaling exponent $p$ that ranges from 1 to 2 concerning computational efficiency. 
An upper bound $v_{\max}$ on cell volumes is chosen to achieve $v_1(D_1/2)\lesssim 0.001 \abs{\mathcal{V}_2}$; see Sec. \ref{sec:sphere-tessellation} and the Supplementary Material.
Figure \ref{fig:multiscale_tiling_initial}(a)-\ref{fig:multiscale_tiling_final}(b) illustrate our multiscale-disk tiling model in the $400$th stage with $p=1.5$, and the resulting coated-disks model at the local-cell packing fraction $\phi=0.25$. 
Figure \ref{fig:multiscale-tessellations}(a) summarizes that covering fraction of the $m$th stage of our model for various values of a scaling exponent $p$. This shows that in finite stages $m$, our model has the uncovered gaps that contribute to the long-wavelength fluctuations and thus becomes nearly hyperuniform, rather than perfectly hyperuniform. 
However, Fig. \ref{fig:multiscale-tessellations}(c) clearly demonstrates as those gaps are filled ($\eta_m \to 1$) with smaller composite disks, the associated fluctuations $\fn{\tilde{\chi}_V ^{(m)}}{k\to0}$ vanish on the order of $(1-\eta_m)^2$. 
This scaling behavior is consistent with our upper bound \eqref{eq:chi_v_coated M} although our prediction is grossly overestimated. 
Thus, from both our theoretical and numerical results, we conclude that the ideal multiscale coated-spheres model with $\eta_m = 1$ will be strongly hyperuniform.

\section{Fabrication of the computational designs}\label{sec:Fabrication}
\begin{figure}[h]
\centering
\includegraphics[width = 0.45\textwidth]{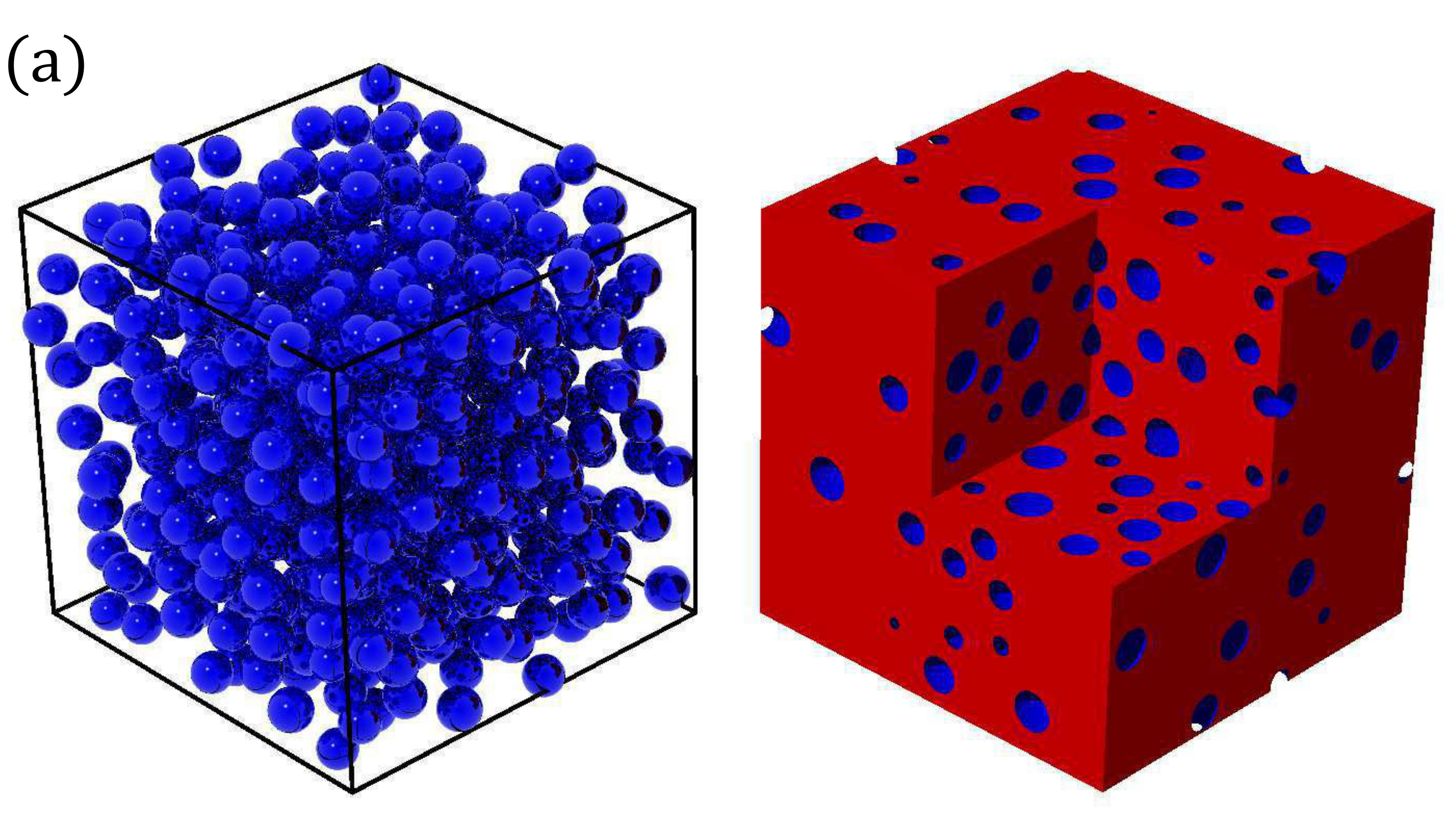}

\includegraphics[width = 0.45\textwidth]{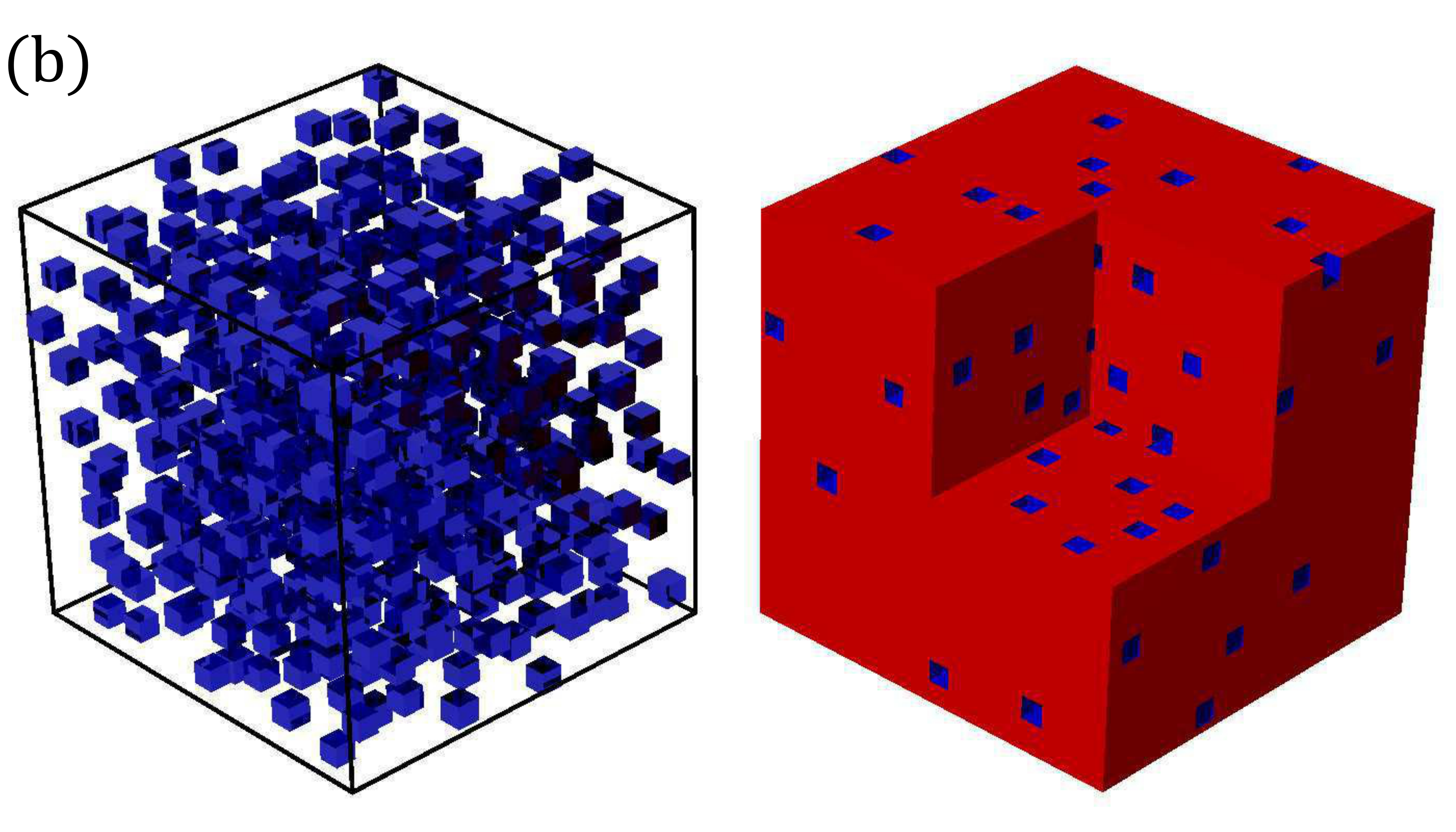}
\caption{Fabrication of designed hyperuniform dispersions and matrices in three dimensions.
(a) A portion of designed dispersion of spherical particles (left) and the corresponding matrix (right) at $\phi=0.23$, which are constructed from a 3D saturated RSA packing.  
(a) A portion of designed dispersion of cubical particles (left) and the corresponding matrix (right).
The hyperuniform matrices with spherical or cubical pores can be fabricated using 3D printing techniques.
\label{fig:fabrications}}
\end{figure}

Importantly, our designed hyperuniform structures obtained from Voronoi tessellations can be easily fabricated via either photolithographic or 3D-printing techniques.
State-of-the-art photolithography fabrication methods are highly suitable for mass production, and capable of creating a 2D pattern on a wafer up to $30$ cm in diameter with $25$ nm in minimal feature size \cite{Zhao2018}.
Since any sharp corners in the designed particle shapes are rounded up to the minimal feature size, it is easier to fabricate circular particles than polygonal ones.
When the minimal feature size is around $1.5~\mu$m, one can easily fabricate our two-dimensional hyperuniform designs with more than one million particles.
For three-dimensional designs, modern 3D-printing techniques \cite{Shirazi2015} can be applied.
Since a printed structure must be topologically connected phase that is mechanically self-supporting, one needs to print the connected matrix phase corresponding to our hyperuniform dispersions \cite{Zerhouni2018} as shown in right panels in Fig. \ref{fig:fabrications}. 
Due to recent advances in 3D-printing techniques, even commercial desktop 3D-printers can achieve  around $100~\mu$m in XY-resolution, $20~\mu$m in Z-resolution \cite{Bhushan2017}, and can print a sample with dimensions $125\times 125 \times 125$ cm$^3$ in 50 hours.
Such devices should be able to print our designs of around 50 million pores when the smallest pore diameter is set around $300\mu$m.

\section{Conclusions}
\label{sec:conclusions} 
In summary, we provide an efficient procedure that is capable of constructing  very large disordered dispersions that are \textit{perfectly} hyperuniform.
Unlike many previous methods that have been used to generate disordered hyperuniform materials, our procedure is simple to implement as it only involves constraining the local-cell packing fraction, which is independent of the rest of a system.
Furthermore, all computations to determine particle volumes can be exactly performed and easily parallelized. 
Our methodology shares some similarities with the ``equal-volume tessellation'' to obtain hyperuniform point patterns \cite{Gabrielli2008} in which the local number density within each cell is identical.
However, our procedure is more versatile in that it can be applied to Voronoi tessellations of disordered point patterns and multiscale-sphere tessellations that we discussed in this paper.

For simplicity, we have focused in this work on constructed dispersions of spherical particles and sketched the proof of their hyperuniformity.
In Ref. \cite{Jaeuk2018_4}, we provide detailed derivations of the spectral densities of hyperuniform dispersions consisting of particles of arbitrary shape and show how any statistically homogeneous progenitor packing satisfies the bounded-cell condition. 
There, we also present additional simulation results for a variety of progenitor sphere packings as well as packings of nonspherical particles. In the case of coated-spheres model, we provide additional mathematical details as well as numerical results using different cell-volume scalings \cite{Jaeuk2018_4}.

In this work, we have shown that local statistics alone generally does not determine hyperuniformity, which although reinforces previous observations (see a recent review \cite{Torquato2018_review} and references therein), is not commonly understood.
While the nonhyperuniform progenitor configurations have no influence on the behavior of the spectral density around the origin, since the constructed dispersion is hyperuniform of class I, it starts to determine its behavior away from the origin at intermediate to large wavenumbers.
Indeed, at intermediate wavenumbers the spectral densities of the progenitor and constructed dispersions collapse onto one another \cite{Jaeuk2018_4} because local statistics, such as Voronoi cell volumes and nearest-neighbor distance distributions, are identical for both systems.
This outcome is consistent with the fact that hyperuniformity is a global property of a system. 

 It should not go unnoticed that many other tessellations can be employed  in our tessellation-based methodology, including generalizations of Voronoi tessellations, such as Laguerre or radical tessellations \cite{Hiroshi1985, Gellatly1982} and Voronoi tessellations in Manhattan distance \cite{Krause_TaxicabGeometry},  Delaunay triangulations \cite{Torquato_RHM},  ``Delaunay-centroidal'' tessellations \cite{Florescu2009, Torquato2018_3}, and disordered isoradial graphs \cite{Deuschel2012} whenever they meet the bounded-cell condition.
Furthermore, our methodology allows one to tune particle shapes and their numbers within each cell while preserving hyperuniformity. This exploration of different tessellations and particle geometries/numbers in each cell enables one to generate an enormous class of hyperuniform dispersions and hence represents fertile ground for future research. 

We have established for the first time that the optimal Hashin-Shtrikman multiscale dispersions are indeed hyperuniform. 
This finding suggests that hyperuniformity is evidently a crucial characteristic to achieve optimality with respect to effective transport and elastic properties \cite{Torquato_RHM, Hashin1962,Torquato2004}. 
Thus, not only do the spheres have to be ``well-separated'' from one another (as is traditionally understood) but the entire dispersion should possess the global property of hyperuniformity to achieve optimal effective physical properties, which heretofore had not been known.  
While it has begun to be shown that some disordered hyperuniform dispersions have desirable and nearly optimal transport, mechanical and optical properties \cite{Leseur2016, Zhang2016, Chen2017},  it will be of great interest to study the physical properties of the enormous class of hyperuniform dispersions that can be designed and tuned by our tessellation-based methodology.
 These computational designs can subsequently be combined with the aforementioned 2D and 3D fabrication techniques to accelerate the discovery of hyperuniform two-phase materials.

\section*{Acknowledgements}
We thank G. Zhang, T. G. Mason, A. B. Hopkins, and M. A. Klatt for very helpful discussions.
This work was partially supported by the U.S. National Science Foundation under Grant No. DMR-1714722.


\begin{thebibliography}{10}
\expandafter\ifx\csname url\endcsname\relax
  \def\url#1{\texttt{#1}}\fi
\expandafter\ifx\csname urlprefix\endcsname\relax\def\urlprefix{URL }\fi
\expandafter\ifx\csname href\endcsname\relax
  \def\href#1#2{#2} \def\path#1{#1}\fi

\bibitem{Torquato2003_hyper}
S.~Torquato, F.~H. Stillinger, Local density fluctuations, hyperuniformity, and
  order metrics, Phys. Rev. E 68~(4) (2003) 041113.
\newblock \href {http://dx.doi.org/10.1103/PhysRevE.68.041113}
  {\path{doi:10.1103/PhysRevE.68.041113}}.

\bibitem{Zachary2009a}
C.~E. Zachary, S.~Torquato, Hyperuniformity in point patterns and two-phase
  random heterogeneous media, J. Stat. Mech: Theory Exp. 2009~(12) (2009)
  P12015.
\newblock \href {http://dx.doi.org/10.1088/1742-5468/2009/12/P12015}
  {\path{doi:10.1088/1742-5468/2009/12/P12015}}.

\bibitem{Torquato2018_review}
S.~Torquato, Hyperuniform {S}tates of {M}atter, Phys. Rep. 745 (2018) 1 -- 95.
\newblock \href {http://dx.doi.org/10.1016/j.physrep.2018.03.001}
  {\path{doi:10.1016/j.physrep.2018.03.001}}.

\bibitem{Florescu2009}
M.~Florescu, S.~Torquato, P.~J. Steinhardt, Designer disordered materials with
  large, complete photonic band gaps, Proc. Natl. Acad. Sci. U.S.A. 106~(49)
  (2009) 20658--20663.
\newblock \href {http://dx.doi.org/10.1073/pnas.0907744106}
  {\path{doi:10.1073/pnas.0907744106}}.

\bibitem{Scheffold2017}
L.~S. Froufe-P\'{e}rez, M.~Engel, J.~J. S\'{a}enz, F.~Scheffold, Band gap
  formation and anderson localization in disordered photonic materials with
  structural correlations, Proc. Natl. Acad. Sci. U.S.A. 114~(36) (2017)
  9570--9574.
\newblock \href {http://dx.doi.org/10.1073/pnas.1705130114}
  {\path{doi:10.1073/pnas.1705130114}}.

\bibitem{Lopez2018}
C.~L\'{o}pez, The true value of disorder, Adv. Opt. Mater. 6~(16) (2018)
  1800439.
\newblock \href {http://dx.doi.org/10.1002/adom.201800439}
  {\path{doi:10.1002/adom.201800439}}.

\bibitem{Zhang2016}
G.~Zhang, F.~Stillinger, S.~Torquato, Transport, geometrical, and topological
  properties of stealthy disordered hyperuniform two-phase systems, J. Chem.
  Phys. 145~(24) (2016) 244109.
\newblock \href {http://dx.doi.org/10.1063/1.4972862}
  {\path{doi:10.1063/1.4972862}}.

\bibitem{Chen2017}
D.~Chen, S.~Torquato, Designing disordered hyperuniform two-phase materials
  with novel physical properties, Acta Mater. 142 (2018) 152--161.
\newblock \href {http://dx.doi.org/10.1016/j.actamat.2017.09.053}
  {\path{doi:10.1016/j.actamat.2017.09.053}}.

\bibitem{Thien2016}
Q.~Le~Thien, D.~McDermott, C.~J.~O. Reichhardt, C.~Reichhardt, Enhanced pinning
  for vortices in hyperuniform pinning arrays and emergent hyperuniform vortex
  configurations with quenched disorder, Phys. Rev. B 96~(9) (2017) 094516.
\newblock \href {http://dx.doi.org/10.1103/PhysRevB.96.094516}
  {\path{doi:10.1103/PhysRevB.96.094516}}.

\bibitem{Leseur2016}
O.~Leseur, R.~Pierrat, R.~Carminati, High-density hyperuniform materials can be
  transparent, Optica 3~(7) (2016) 763--767.
\newblock \href {http://dx.doi.org/10.1364/OPTICA.3.000763}
  {\path{doi:10.1364/OPTICA.3.000763}}.

\bibitem{Brinker_sol-gel}
C.~J. Brinker, G.~W. Scherer, Sol-gel science: {T}he physics and chemistry of
  sol-gel processing, Academic press, 2013.

\bibitem{Sahimi_HM1}
M.~Sahimi, Heterogeneous {M}aterials {I}: {L}inear {T}ransport and {O}ptical
  {P}roperties, Vol.~22, Springer-Verlag, New York, 2003.

\bibitem{Torquato_RHM}
S.~Torquato, Random {H}eterogeneous {M}aterials: {M}icrostructure and
  {M}acroscopic {P}roperties, Interdisciplinary Applied Mathematics, Springer
  Science \& Business Media, 2002.

\bibitem{Patel2016}
B.~Patel, T.~I. Zohdi, Numerical estimation of effective electromagnetic
  properties for design of particulate composites, Mater. Des. 94 (2016)
  546--553.
\newblock \href {http://dx.doi.org/10.1016/j.matdes.2016.01.015}
  {\path{doi:10.1016/j.matdes.2016.01.015}}.

\bibitem{Mickel2013}
W.~Mickel, S.~C. Kapfer, G.~E. Schr\"{o}der-Turk, K.~Mecke, Shortcomings of the
  bond orientational order parameters for the analysis of disordered
  particulate matter, J. Chem. Phys. 138~(4) (2013) 044501.
\newblock \href {http://dx.doi.org/10.1063/1.4774084}
  {\path{doi:10.1063/1.4774084}}.

\bibitem{Debye1949}
P.~Debye, A.~M. Bueche, Scattering by an inhomogeneous solid, J. Appl. Phys.
  20~(6) (1949) 518--525.
\newblock \href {http://dx.doi.org/10.1063/1.1698419}
  {\path{doi:10.1063/1.1698419}}.

\bibitem{Uche2004}
O.~U. Uche, F.~H. Stillinger, S.~Torquato, Constraints on collective density
  variables: {T}wo dimensions, Phys. Rev. E 70~(4) (2004) 046122.
\newblock \href {http://dx.doi.org/10.1103/PhysRevE.70.046122}
  {\path{doi:10.1103/PhysRevE.70.046122}}.

\bibitem{Torquato2015_stealthy}
S.~Torquato, G.~Zhang, F.~Stillinger, Ensemble theory for stealthy hyperuniform
  disordered ground states, Phys. Rev. X 5~(2) (2015) 021020.
\newblock \href {http://dx.doi.org/10.1103/PhysRevX.5.021020}
  {\path{doi:10.1103/PhysRevX.5.021020}}.

\bibitem{Zhang2015}
G.~Zhang, F.~H. Stillinger, S.~Torquato, Ground states of stealthy hyperuniform
  potentials: {I}. {E}ntropically favored configurations, Phys. Rev. E 92~(2)
  (2015) 022119.
\newblock \href {http://dx.doi.org/10.1103/PhysRevE.92.022119}
  {\path{doi:10.1103/PhysRevE.92.022119}}.

\bibitem{Dyson1962}
F.~J. Dyson, Statistical theory of the energy levels of complex systems. {I},
  J. Math. Phys. 3~(1) (1962) 140--156.
\newblock \href {http://dx.doi.org/10.1063/1.1703773}
  {\path{doi:10.1063/1.1703773}}.

\bibitem{Jancovici1981}
B.~Jancovici, Exact results for the two-dimensional one-component plasma, Phys.
  Rev. Lett. 46~(6) (1981) 386.
\newblock \href {http://dx.doi.org/10.1103/PhysRevLett.46.386}
  {\path{doi:10.1103/PhysRevLett.46.386}}.

\bibitem{Duyu2018_2}
D.~Chen, E.~Lomba, S.~Torquato, Binary mixtures of charged colloids: {A}
  potential route to synthesize disordered hyperuniform materials, Phys. Chem.
  Chem. Phys. 20~(26) (2018) 17557--17562.
\newblock \href {http://dx.doi.org/10.1039/c8cp02616e}
  {\path{doi:10.1039/c8cp02616e}}.

\bibitem{Feynman1956}
R.~Feynman, M.~Cohen, Energy spectrum of the excitations in liquid helium,
  Phys. Rev. 102~(5) (1956) 1189.
\newblock \href {http://dx.doi.org/10.1103/PhysRev.102.1189}
  {\path{doi:10.1103/PhysRev.102.1189}}.

\bibitem{Torquato2008}
S.~Torquato, A.~Scardicchio, C.~E. Zachary, Point processes in arbitrary
  dimension from fermionic gases, random matrix theory, and number theory, J.
  Stat. Mech: Theory Exp. 2008~(11) (2008) P11019.
\newblock \href {http://dx.doi.org/10.1088/1742-5468/2008/11/P11019}
  {\path{doi:10.1088/1742-5468/2008/11/P11019}}.

\bibitem{Scardicchio2009}
A.~Scardicchio, C.~E. Zachary, S.~Torquato, Statistical properties of
  determinantal point processes in high-dimensional euclidean spaces, Phys.
  Rev. E 79~(4) (2009) 041108.
\newblock \href {http://dx.doi.org/10.1103/PhysRevE.79.041108}
  {\path{doi:10.1103/PhysRevE.79.041108}}.

\bibitem{Donev2005}
A.~Donev, F.~H. Stillinger, S.~Torquato, Unexpected density fluctuations in
  jammed disordered sphere packings, Phys. Rev. Lett. 95~(9) (2005) 090604.
\newblock \href {http://dx.doi.org/10.1103/PhysRevLett.95.090604}
  {\path{doi:10.1103/PhysRevLett.95.090604}}.

\bibitem{Kurita2011}
R.~Kurita, E.~R. Weeks, Incompressibility of polydisperse random-close-packed
  colloidal particles, Phys. Rev. E 84~(R) (2011) 030401.
\newblock \href {http://dx.doi.org/10.1103/PhysRevE.84.030401}
  {\path{doi:10.1103/PhysRevE.84.030401}}.

\bibitem{Jiao2011}
Y.~Jiao, S.~Torquato, Maximally random jammed packings of {P}latonic solids:
  {H}yperuniform long-range correlations and isostaticity, Phys. Rev. E 84~(4)
  (2011) 041309.
\newblock \href {http://dx.doi.org/10.1103/PhysRevE.84.041309}
  {\path{doi:10.1103/PhysRevE.84.041309}}.

\bibitem{Hexner2015}
D.~Hexner, D.~Levine, Hyperuniformity of {C}ritical {A}bsorbing {S}tates, Phys.
  Rev. Lett. 114~(11) (2015) 110602.
\newblock \href {http://dx.doi.org/10.1103/PhysRevLett.114.110602}
  {\path{doi:10.1103/PhysRevLett.114.110602}}.

\bibitem{Weijs2015}
J.~H. Weijs, R.~Jeanneret, R.~Dreyfus, D.~Bartolo, Emergent hyperuniformity in
  periodically driven emulsions, Phys. Rev. Lett. 115~(10) (2015) 108301.
\newblock \href {http://dx.doi.org/10.1103/PhysRevLett.115.108301}
  {\path{doi:10.1103/PhysRevLett.115.108301}}.

\bibitem{Garcia-Millan2018}
R.~Garcia-{M}illan, G.~Pruessner, L.~Pickering, K.~Christensen, Correlations
  and hyperuniformity in the avalanche size of the oslo model, Europhys. Lett.
  122~(5) (2018) 50003.
\newblock \href {http://dx.doi.org/10.1209/0295-5075/122/50003}
  {\path{doi:10.1209/0295-5075/122/50003}}.

\bibitem{Battista2018}
D.~Di~Battista, D.~Ancora, G.~Zacharakis, G.~Ruocco, M.~Leonetti,
  Hyperuniformity in amorphous speckle patterns, Opt. Express 26~(12) (2018)
  15594--15608.
\newblock \href {http://dx.doi.org/10.1364/OE.26.015594}
  {\path{doi:10.1364/OE.26.015594}}.

\bibitem{Torquato2018_prime2}
S.~Torquato, G.~Zhang, M.~de~Courcy-Ireland, Uncovering multiscale order in the
  prime numbers via scattering, J. Stat. Mech: Theory Exp. 2018~(9) (2018)
  093401.
\newblock \href {https://doi.org/10.1088/1742-5468/aad6be}
  {\path{doi:10.1088/1742-5468/aad6be}}.
\bibitem{Montgomery1973}
H.~L. Montgomery, The pair correlation of zeros of the zeta function, in: Proc.
  Symp. Pure Math, Vol.~24, 1973, pp. 181--193.

\bibitem{Sodin2006}
M.~Sodin, B.~Tsirelson, Random complex zeroes, {II}. {P}erturbed lattice,
  Israel J. Math. 152~(1) (2006) 105--124.
\newblock \href {http://dx.doi.org/10.1007/Bf02771978}
  {\path{doi:10.1007/Bf02771978}}.

\bibitem{Noh2010}
H.~Noh, S.~F. Liew, V.~Saranathan, S.~G.~J. Mochrie, R.~O. Prum, E.~R.
  Dufresne, H.~Cao, How noniridescent colors are generated by quasi-ordered
  structures of bird feathers, Adv. Mater. 22~(26-27) (2010) 2871--2880.
\newblock \href {http://dx.doi.org/10.1002/adma.200903699}
  {\path{doi:10.1002/adma.200903699}}.

\bibitem{Jiao2014_chickenEyes}
Y.~Jiao, T.~Lau, H.~Hatzikirou, M.~Meyer-Hermann, C.~C. Joseph, S.~Torquato,
  Avian photoreceptor patterns represent a disordered hyperuniform solution to
  a multiscale packing problem, Phys. Rev. E 89~(2) (2014) 022721.
\newblock \href {http://dx.doi.org/10.1103/PhysRevE.89.022721}
  {\path{doi:10.1103/PhysRevE.89.022721}}.

\bibitem{Mayer2015}
A.~Mayer, V.~Balasubramanian, T.~Mora, A.~M. Walczak, How a well-adapted immune
  system is organized, Proc. Natl. Acad. Sci. U.S.A. 112~(19) (2015)
  5950--5955.
\newblock \href {http://dx.doi.org/10.1073/pnas.1421827112}
  {\path{doi:10.1073/pnas.1421827112}}.

\bibitem{Kwon2017}
S.~Kwon, J.~M. Kim, Hyperuniformity of initial conditions and critical decay of
  a diffusive epidemic process belonging to the manna class, Phys. Rev. E
  96~(1) (2017) 012146.
\newblock \href {http://dx.doi.org/10.1103/PhysRevE.96.012146}
  {\path{doi:10.1103/PhysRevE.96.012146}}.

\bibitem{Atkinson2016}
S.~Atkinson, G.~Zhang, A.~B. Hopkins, S.~Torquato, Critical slowing down and
  hyperuniformity on approach to jamming, Phys. Rev. E 94~(1-1) (2016) 012902.
\newblock \href {http://dx.doi.org/10.1103/PhysRevE.94.012902}
  {\path{doi:10.1103/PhysRevE.94.012902}}.

\bibitem{Hejna2013}
M.~Hejna, P.~J. Steinhardt, S.~Torquato, Nearly hyperuniform network models of
  amorphous silicon, Phys. Rev. B 87~(24) (2013) 245204.
\newblock \href {http://dx.doi.org/10.1103/PhysRevB.87.245204}
  {\path{doi:10.1103/PhysRevB.87.245204}}.

\bibitem{Florescu2010}
M.~Florescu, S.~Torquato, P.~Steinhardt, Effects of random link removal on the
  photonic band gaps of honeycomb networks, Appl. Phys. Lett. 97~(20) (2010)
  201103.
\newblock \href {http://dx.doi.org/10.1063/1.3505322}
  {\path{doi:10.1063/1.3505322}}.

\bibitem{Ricouvier2017}
J.~Ricouvier, R.~Pierrat, R.~Carminati, P.~Tabeling, P.~Yazhgur, Optimizing
  hyperuniformity in self-assembled bidisperse emulsions, Phys. Rev. Lett.
  119~(20) (2017) 208001.
\newblock \href {http://dx.doi.org/10.1103/PhysRevLett.119.208001}
  {\path{doi:10.1103/PhysRevLett.119.208001}}.

\bibitem{Zito2015}
G.~Zito, G.~Rusciano, G.~Pesce, A.~Malafronte, R.~Di~Girolamo, G.~Ausanio,
  A.~Vecchione, A.~Sasso, Nanoscale engineering of two-dimensional disordered
  hyperuniform block-copolymer assemblies, Phys. Rev. E 92~(5) (2015) 050601.
\newblock \href {http://dx.doi.org/10.1103/PhysRevE.92.050601}
  {\path{doi:10.1103/PhysRevE.92.050601}}.

\bibitem{Zachary2011_3}
C.~E. Zachary, Y.~Jiao, S.~Torquato, Hyperuniform long-range correlations are a
  signature of disordered jammed hard-particle packings, Phys. Rev. Lett.
  106~(17) (2011) 178001.
\newblock \href {http://dx.doi.org/10.1103/PhysRevLett.106.178001}
  {\path{doi:10.1103/PhysRevLett.106.178001}}.

\bibitem{Lomba2017}
E.~Lomba, J.~J. Weis, S.~Torquato, Disordered hyperuniformity in two-component
  nonadditive hard-disk plasmas, Phys. Rev. E 96~(6-1) (2017) 062126.
\newblock \href {http://dx.doi.org/10.1103/PhysRevE.96.062126}
  {\path{doi:10.1103/PhysRevE.96.062126}}.

\bibitem{TJ_algorithm}
S.~Torquato, Y.~Jiao, Robust algorithm to generate a diverse class of dense
  disordered and ordered sphere packings via linear programming, Phys. Rev. E
  82~(6) (2010) 061302.
\newblock \href {http://dx.doi.org/10.1103/PhysRevE.82.061302}
  {\path{doi:10.1103/PhysRevE.82.061302}}.

\bibitem{LS_algorithm}
B.~D. Lubachevsky, F.~H. Stillinger, Geometric properties of random disk
  packings, J. Stat. Phys. 60~(5-6) (1990) 561--583.
\newblock \href {http://dx.doi.org/10.1007/Bf01025983}
  {\path{doi:10.1007/Bf01025983}}.

\bibitem{Jaeuk2018}
J.~Kim, S.~Torquato, Effect of imperfections on the hyperuniformity of
  many-body systems, Phys. Rev. B 97~(5) (2018) 054105.
\newblock \href {http://dx.doi.org/10.1103/PhysRevB.97.054105}
  {\path{doi:10.1103/PhysRevB.97.054105}}.

\bibitem{Torquato2018_packing}
S.~Torquato, Perspective: Basic understanding of condensed phases of matter via
  packing models, J. Chem. Phys. 149~(2) (2018) 020901.
\newblock \href {http://dx.doi.org/10.1063/1.5036657}
  {\path{doi:10.1063/1.5036657}}.

\bibitem{Zhao2018}
K.~Zhao, T.~G. Mason, Assembly of colloidal particles in solution, Rep. Prog.
  Phys. 80 (2018) 126601.
\newblock \href {http://dx.doi.org/10.1088/1361-6633/aad1a7}
  {\path{doi:10.1088/1361-6633/aad1a7}}.

\bibitem{Wong2012}
K.~V. Wong, A.~Hernandez, A review of additive manufacturing, ISRN mech eng
  2012 (2012) 1--10.
\newblock \href {http://dx.doi.org/10.5402/2012/208760}
  {\path{doi:10.5402/2012/208760}}.

\bibitem{Tumbleston2015}
J.~R. Tumbleston, D.~Shirvanyants, N.~Ermoshkin, R.~Janusziewicz, A.~R.
  Johnson, D.~Kelly, K.~Chen, R.~Pinschmidt, J.~P. Rolland, A.~Ermoshkin, E.~T.
  Samulski, J.~M. DeSimone, Continuous liquid interface production of 3{D}
  objects, Science 347~(6228) (2015) 1349--1352.
\newblock \href {http://dx.doi.org/10.1126/science.aaa2397}
  {\path{doi:10.1126/science.aaa2397}}.

\bibitem{Shirazi2015}
S.~F. Shirazi, S.~Gharehkhani, M.~Mehrali, H.~Yarmand, H.~S. Metselaar,
  N.~Adib~Kadri, N.~A. Osman, A review on powder-based additive manufacturing
  for tissue engineering: {S}elective laser sintering and inkjet {3D} printing,
  Sci. Technol. Adv. Mater. 16~(3) (2015) 033502.
\newblock \href {http://dx.doi.org/10.1088/1468-6996/16/3/033502}
  {\path{doi:10.1088/1468-6996/16/3/033502}}.

\bibitem{Torquato1999_3}
S.~Torquato, Exact conditions on physically realizable correlation functions of
  random media, J. Chem. Phys. 111~(19) (1999) 8832--8837.
\newblock \href {http://dx.doi.org/10.1063/1.480255}
  {\path{doi:10.1063/1.480255}}.

\bibitem{Torquato2016_gen}
S.~Torquato, Hyperuniformity and its generalizations, Phys. Rev. E 94~(2)
  (2016) 022122.
\newblock \href {http://dx.doi.org/10.1103/PhysRevE.94.022122}
  {\path{doi:10.1103/PhysRevE.94.022122}}.

\bibitem{Durian2017}
D.~J. Durian, Hyperuniformity disorder length spectroscopy for extended
  particles, Phys. Rev. E 96~(3) (2017) 032910.
\newblock \href {http://dx.doi.org/10.1103/PhysRevE.96.032910}
  {\path{doi:10.1103/PhysRevE.96.032910}}.

\bibitem{Wisdom1966}
B.~Widom, Random sequential addition of hard spheres to a volume, J. Chem.
  Phys. 44~(10) (1966) 3888--3894.
\newblock \href {http://dx.doi.org/10.1063/1.1726548}
  {\path{doi:10.1063/1.1726548}}.

\bibitem{Zhang2013}
G.~Zhang, S.~Torquato, Precise algorithm to generate random sequential addition
  of hard hyperspheres at saturation, Phys. Rev. E 88~(5) (2013) 053312.
\newblock \href {http://dx.doi.org/10.1103/PhysRevE.88.053312}
  {\path{doi:10.1103/PhysRevE.88.053312}}.

\bibitem{Gellatly1982}
B.~J. Gellatly, J.~L. Finney, Characterisation of models of multicomponent
  amorphous metals: The radical alternative to the {V}oronoi polyhedron, J.
  Non-Cryst. Solids 50~(3) (1982) 313--329.
\newblock \href {http://dx.doi.org/10.1016/0022-3093(82)90093-X}
  {\path{doi:10.1016/0022-3093(82)90093-X}}.

\bibitem{Hiroshi1985}
H.~Imai, M.~Iri, K.~Murota, Voronoi {D}iagram in the {L}aguerre {G}eometry and
  {I}ts {A}pplications, SIAM J. Comput. 14~(1) (1985) 93--105.
\newblock \href {http://dx.doi.org/10.1137/0214006}
  {\path{doi:10.1137/0214006}}.

\bibitem{Krause_TaxicabGeometry}
E.~F. Krause, Taxicab geometry: An adventure in non-Euclidean geometry, Courier
  Corporation, 1986.

\bibitem{voro++}
C.~H. Rycroft, {VORO++}: a three-dimensional {V}oronoi cell library in {C}++,
  Chaos 19~(4) (2009) 041111.
\newblock \href {http://dx.doi.org/10.1063/1.3215722}
  {\path{doi:10.1063/1.3215722}}.

\bibitem{Shier2013}
J.~Shier, P.~Bourke, An algorithm for random fractal filling of space, Comput.
  Graphics Forum 32~(8) (2013) 89--97.
\newblock \href {http://dx.doi.org/10.1111/cgf.12163}
  {\path{doi:10.1111/cgf.12163}}.

\bibitem{Jaeuk2018_4}
J.~Kim, S.~Torquato, Methodology to Construct Large Realizations of Perfectly Hyperuniform Disordered Packings (2019) in preparation.
\newblock \href{https://arxiv.org/abs/1901.10006}
  {\path{arXiv:1901.10006}}

\bibitem{Torquato1991}
S.~Torquato, M.~Avellaneda, Diffusion and reaction in heterogeneous media: Pore
  size distribution, relaxation times, and mean survival time, J. Chem. Phys.
  95~(9) (1991) 6477--6489.
\newblock \href {http://dx.doi.org/10.1063/1.461519}
  {\path{doi:10.1063/1.461519}}.

\bibitem{Kampf2015}
J.~Kampf, E.~Spodarev, A functional central limit theorem for integrals of
  stationary mixing random fields, Theory Probab. Its Appl. 63~(1) (2017)
  135--150.
\newblock \href {http://dx.doi.org/10.1137/S0040585X97T988952}
  {\path{doi:10.1137/S0040585X97T988952}}.

\bibitem{Hashin1962}
Z.~Hashin, The elastic moduli of heterogeneous materials, J. Appl. Mech. 29~(1)
  (1962) 143--150.
\newblock \href {http://dx.doi.org/10.1115/1.3636446}
  {\path{doi:10.1115/1.3636446}}.

\bibitem{Torquato2004}
S.~Torquato, D.~Pham, Optimal bounds on the trapping constant and permeability
  of porous media, Phys. Rev. Lett. 92~(25) (2004) 255505.
\newblock \href {http://dx.doi.org/10.1103/PhysRevLett.92.255505}
  {\path{doi:10.1103/PhysRevLett.92.255505}}.

\bibitem{Torquato2018_3}
S.~Torquato, D.~Chen, Multifunctional hyperuniform cellular networks:
  optimality, anisotropy and disorder, Multifunct. Mater. 1~(1) (2018) 015001.
\newblock \href {http://dx.doi.org/10.1088/2399-7532/aaca91}
  {\path{doi:10.1088/2399-7532/aaca91}}.

\bibitem{Torquato2018_5}
S.~Torquato, D.~Chen, Multifunctionality of particulate composites via
  cross-property maps, Phys. Rev. Mater. 2~(9) (2018) 095603.
\newblock \href {http://dx.doi.org/10.1103/PhysRevMaterials.2.095603}
  {\path{doi:10.1103/PhysRevMaterials.2.095603}}.

\bibitem{Zerhouni2018}
O.~Zerhouni, M.-G. Tarantino, K.~Danas, F.~Hong, Influence of the internal
  geometry on the elastic properties of materials using 3{D} printing of
  computer-generated random microstructures, Society of Exploration
  Geophysicists, 2018, pp. 3713--3718.
\newblock \href {http://dx.doi.org/10.1190/segam2018-2998182.1}
  {\path{doi:10.1190/segam2018-2998182.1}}.

\bibitem{Bhushan2017}
B.~Bhushan, M.~Caspers, An overview of additive manufacturing (3d printing) for
  microfabrication, Microsyst. Technol. 23~(4) (2017) 1117--1124.
\newblock \href {http://dx.doi.org/10.1007/s00542-017-3342-8}
  {\path{doi:10.1007/s00542-017-3342-8}}.

\bibitem{Gabrielli2008}
A.~Gabrielli, M.~Joyce, S.~Torquato, Tilings of space and superhomogeneous
  point processes, Phys. Rev. E 77~(3) (2008) 031125.
\newblock \href {http://dx.doi.org/10.1103/PhysRevE.77.031125}
  {\path{doi:10.1103/PhysRevE.77.031125}}.

\bibitem{Deuschel2012}
C.~Boutillier, B.~de~Tili\`{e}re, Statistical mechanics on isoradial graphs,
  in: J.-D. Deuschel, B.~Gentz, W.~K\"{o}nig, M.~von Renesse, M.~Scheutzow,
  U.~Schmock (Eds.), Probability in Complex Physical Systems, Springer Berlin
  Heidelberg, 2012, pp. 491--512.

\end{thebibliography}

\end{document}